\documentclass[ejsv2]{imsart}
\pubyear{2025}
\volume{0}
\firstpage{0}
\lastpage{0}

\usepackage{amsthm,amsmath,amsfonts,amssymb,mathrsfs}
\usepackage{bbm}
\usepackage{physics}
\usepackage{eucal}
\usepackage{natbib}
\usepackage{xcolor}
\RequirePackage[colorlinks,citecolor=blue,urlcolor=blue]{hyperref}
\RequirePackage{graphicx}

\arxiv{2010.00000}
\startlocaldefs
\theoremstyle{plain}

\newtheorem{theorem}{Theorem}[section]
\newtheorem{lemma}{Lemma}
\def\bI{\boldsymbol{I}}
\def\bJ{\boldsymbol{J}}

\begin{document}
\begin{frontmatter}
\title{Debiased inference in error-in-variable problems with non-Gaussian measurement error}
\runtitle{Debiased inference in error-in-variable problems}

\begin{aug}

\author{\fnms{Nicholas W.} \snm{Woolsey} and \fnms{Xianzheng}  \snm{Huang}  }
\address{Department of Statistics, University of South Carolina\\ \printead{e1}}
\author{\ead[label=e1]{nwoolsey@email.sc.edu; huang@stat.sc.edu}}

\runauthor{Woolsey \and Huang}
\end{aug}

\begin{abstract}
We consider drawing statistical inferences based on data subject to non-Gaussian measurement error. Unlike most existing methods developed under the assumption of Gaussian measurement error, the proposed strategy exploits hypercomplex numbers to reduce bias in naive estimation that ignores non-Gaussian measurement error. We apply this new method to several widely applicable parametric regression models with error-prone covariates, and kernel density estimation using error-contaminated data. The efficacy of this method in bias reduction is demonstrated in simulation studies and a real-life application in sports analytics.
\end{abstract}



\end{frontmatter}

\section{Introduction}
Data subject to measurement error due to imprecise measuring devices or human error are ubiquitous in many applications. There is a rich collection of literature on the study of effects of measurement error on statistical inferences and strategies for addressing complications caused by measurement error. \cite{Fuller}, \cite{carroll2006measurement}, \citet{buonaccorsi2010measurement}, \citet{yi2017statistical} and \cite{grace2021handbook}, among several other books on this subject, provide comprehensive reviews of error-in-variable problems. The history of research on these problems has mainly focused on Gaussian measurement errors.  Several landmark methods including the corrected score method \citep{nakamura, StefNovick2002}, the conditional score method  \citep{condscore}, and the method of simulation-and-extrapolation \citep[SIMEX,][]{simex} were all first developed under the assumption of Gaussian measurement error. However, non-Gaussian measurement errors, such as skewed or heavy-tailed errors, are common in fields including finance and actuarial science \citep{finance}, political science \citep{political}, and environmental science \citep{wind}. There have been some attempts to account for non-Gaussian measurement errors in limited model settings, primarily in linear regression. Most existing methods along this theme assume some specific type of error such as skew-normal \citep{skewnorm}, or assume a known distribution for the unobserved error-prone variable \citep{tthing}.

In this study, we propose a new strategy exploiting hypercomplex numbers to account for measurement errors not limited to Gaussian errors in a wide range of inference problems. 
The idea of utilizing hypercomplex quantities to effectively mitigate the adverse effects of measurement error is developed in Section~\ref{sec:method}. Section~\ref{sec:implement} addresses some unique issues in implementing the proposed method involving hypercomplex numbers. We apply the new method in four parametric regression settings under the framework of M-estimation in Section~\ref{sec:Mest}, and then apply it to the problem of nonparametric density estimation in Section~\ref{sec:density}. Empirical evidence for the method's efficacy in comparison with several competing methods is presented in Section~\ref{sec:simu} using synthetic data. Section~\ref{sec:real} presents a real-life data application of these considered methods. We conclude the paper with a brief discussion on the future research agenda in Section~\ref{sec:disc}. 

\section{Methodology}
\label{sec:method}
\subsection{Correcting for Gaussian  measurement error using complex numbers}
Let $X$ be a random variable of interest that cannot be observed due to error contamination in a study. Suppose that the error-contaminated surrogate is $W$, relating to $X$ via 
\begin{equation}
    W=X+U, \label{eq:addme}
\end{equation} 
where $U$ is the measurement error supported on the real line $\mathbb{R}$ independent of $X$ and other random quantities in the study, such as the response $Y$ and the random error in a regression model. In other words, $U$ is an additive nondifferential measurement error \citep[][Section 2.5]{carroll2006measurement}. A popular recipe for developing methods to draw inferences based on error-contaminated data consists of two main steps. Firstly, one adopts an inference method had $X$ been observed. This usually amounts to choosing a statistic such as an estimator for a target parameter, a score function used in M-estimation \citep{stefanski2002calculus}, or an objective function such as a log-likelihood function or a loss function in the absence of measurement error. Let $f(X)$ generically refer to such a choice evaluated at one data point, with its potential dependence on parameters and other random variables suppressed. Secondly, acknowledging that $X$ is not observed but $W$ is, one seeks to ``recover'' $f(X)$ by constructing a sensible estimator for it based on $W$ or replicate error-contaminated measures of $X$. Because $f(W)$ typically does not serve this purpose but is a naive and often biased estimator for $f(X)$, the  second step sparks many proposals for estimating $f(X)$. These proposals, as well as  ours, focus on functional measurement error models where no distributional assumption is imposed on $X$, unlike structural models that include an assumed distribution of $X$ \citep[][Section 2.1]{carroll2006measurement}. This prolific two-step recipe has led to kernel density estimators based on data subject to measurement error \citep{stefanski1990deconvolving, huang2020conditional}, local polynomial regression with error-in-covariate \citep{delaigle2009design, zhou2016nonparametric}, the corrected score method applied to various (semi-)parametric models \citep{nakamura, StefNovick2002, song2005corrected, wang2006corrected, wu2015smoothed}, and the corrected loss/utility method in different regression settings with error-prone covariates \citep{augustin2004exact, wang2012corrected, li2019linear, liu2024parametric}.  

Along this line of development, \citet{stefanski1989unbiased} introduced complex errors to address the problem in the second step, where it is assumed that (i) $f(\cdot)$ is a known entire function, and (ii) the additive nondifferential error $U\sim N(0, \sigma_u^2)$, with the variance $\sigma_u^2$ known. The author then phrased the problem of estimating $f(X)$ based on $W$ as estimating an entire function of a normal mean. The solution to the problem is inspired by the intriguing result that we reiterate in Theorem~\ref{thm:unbiased} for its high relevance in our methodological development. 
 \begin{theorem}
 \label{thm:unbiased}
Under the following two conditions: (i) $f(\cdot)$ is an entire function over the complex plane; (ii) $U \sim N(0,\sigma_u^2)$ in \eqref{eq:addme}, one has
$E\{f(X + U +i V)|X\} = f(X)$, 
where $i=(-1)^{1/2}$ is the imaginary unit, and $V \sim N(0, \sigma_u^2)$, which is independent of $U$ and $X$.
 \end{theorem}
The expectation of a complex random variable is a complex number of which the real and imaginary parts are equal to the expectations of the real and imaginary parts of the random variable, respectively. According to Theorem~\ref{thm:unbiased}, under conditions (i) and (ii), $f(W+iV)$ is an unbiased estimator for $f(X)$. Essentially, by adding the complex random noise $iV$ to $W$, $f(W+iV)$ corrects the naive estimator $f(W)$ for measurement error, as if it removed measurement error in the naive estimator to recover $f(X)$. This correction stems from the fact that $U+iV$ is a circularly symmetric random variable, which belongs to a type of random variables widely studied in the field of signal processing \citep{picinbono, schreier2010statistical,  adali2011complex}. 

A complex random variable $C$ is circularly symmetric if $e^{it}C$ and $C$ follow the same distribution for any real constant $t$. This definition leads to Lemma~\ref{lem:zeromom}, proved in Appendix~\ref{app:provelem1}.
 \begin{lemma}
 \label{lem:zeromom}
If a complex random variable $C$ is circularly symmetric and has moments of all orders, then
$E(C^\ell)=0$, for any $\ell$ in the set of natural numbers $\mathbb{N}$. 
\end{lemma}
Now consider $C=U+iV$ introduced in Theorem~\ref{thm:unbiased}. By construction, $C$ follows a central complex Gaussian distribution with the probability density function $p_{\hbox {\tiny $C$}}(c)=(2 \pi \sigma_u^2)^{-1}\exp\{-\bar c c /(2\sigma_u^2)\}$, where $\bar c$ is the conjugate of $c$  \citep{goodman1963statistical}. Because $p_{\hbox {\tiny $C$}}(c)$ depends on $c$ only via its magnitude $|c|=(\bar c c)^{1/2}$, rotating $C$ does not change the distribution as the magnitude is invariant under rotation. Hence, $e^{it}C$, as a rotation transformation of $C$, is identically distributed as $C$ for any $t\in \mathbb{R}$, and thus the so-constructed $C$ is circularly symmetric. Returning to the unbiased estimator of $f(X)$ revealed in Theorem~\ref{thm:unbiased}, we have, using a Taylor expansion around $X$, 
\begin{align}
   &\ f(W+iV) \nonumber \\
   = &\ f(X+C)=f(X)+\sum_{\ell=1}^\infty \frac{f^{(\ell)}(X)}{\ell!} C^\ell, \label{eq:taylor}
\end{align}
under the entirety assumption (i) for $f(\cdot) $ that guarantees existence of derivatives of all orders, $\{ f^{(\ell)}(\cdot),  \, \ell \in \mathbb{N}\}$.Taking the expectation of \eqref{eq:taylor} conditional on $X$ gives $E\{f(X+C)|X\}=f(X)$ since a complex Gaussian possesses moments of all orders and, by Lemma~\ref{lem:zeromom}, $E(C^\ell)=0$, for $\ell\in \mathbb{N}$. This concludes Theorem~\ref{thm:unbiased}.

Denote by $\Re(c)$ the real part of a complex number $c$. 
Because, by Theorem~\ref{thm:unbiased}, the imaginary part of $f(W+iV)$ has a mean of zero, one may use $\Re\{f(W+iV)\}$ as the final estimator for $f(X)$. The Monte Carlo corrected score method \citep{StefNovick2002} amounts to using the {\underline M}onte {\underline C}arlo average, $\hat f_{\text{mc}}(X)=\sum_{m=1}^M \Re\{f(W+iV_m)\}/M$, as an estimator of $f(X)$ based on a random sample $(V_1, \ldots, V_M)$ from $N(0, \sigma_u^2)$.  

\subsection{Correcting for non-Gaussian measurement error using complex numbers}
\label{sec:fcc}
Keeping assumption (i), if we relax assumption (ii) by allowing non-Gaussian $U$, then terms in the infinite sum in \eqref{eq:taylor} may not reduce to zero for all $\ell \in \mathbb{N}$ when taking expectations. This is because it is practically infeasible to construct an independent complex random variable that adds to the non-Gaussian $U$ to produce a circularly symmetric $C$ with all moments existing as accomplished above. Indeed, a circularly symmetric distribution that is not complex Gaussian may not even possess a density \citep{ollila2012complex}, hinting at the infeasibility of generating a circularly symmetric $C$ with all moments well-defined when $U$ is non-Gaussian. With circular symmetry practically out of reach in this case, a more fruitful direction to pursue is to formulate $C$ so that some terms of the Taylor series in \eqref{eq:taylor} have mean zero, in the hope of reducing,  although not fully eliminating, the bias of $f(W)$. As a starter, we strive to have the first two terms of the infinite sum in \eqref{eq:taylor} reduce to zero in expectation when $U$ follows some distribution with variance $\sigma_u^2$ but not necessarily mean-zero or Gaussian. To this end, we turn to {\it proper} complex random variables, a notion first introduced in  \citet{neeser1993proper}.

A complex random variable $C$ is said to be proper if its pseudovariance vanishes \citep[][Definition 1]{neeser1993proper}, where the pseudovariance is  defined as $\text{pvar}(C) = E[\{C-E(C)\}^{2}]$, in contrast to the variance $\text{var}(C) =E\{|C-E(C)|^2\}$. 
We establish Theorem~\ref{thm:proper} in Appendix~\ref{app:provethm2.2} that provides a way to convert a generic real $U$ to a proper complex random variable.
 \begin{theorem}
 \label{thm:proper}
If the first two moments of $U$ exist, then $C = U-E(U)+i\{V-E(V)\}$ is proper, where $V$ is independent of $U$ and follows the same distribution as $U$. 
 \end{theorem}
Clearly, the proper $C$ introduced in Theorem~\ref{thm:proper} includes $U+iV$ appearing in Theorem~\ref{thm:unbiased} as a special case under condition (ii) stating that $U\sim N(0, \sigma_u^2)$.  By Theorem~\ref{thm:proper}, if we use in \eqref{eq:taylor} $C = U-E(U)+i\{V-E(V)\}$ that follows some central complex distribution not necessarily complex Gaussian, then the first two terms of the infinite sum in \eqref{eq:taylor} vanish when taking the expectation. This suggests that $f(X+C)=f(W-E(U)+i\{V-E(V)\})$, or keeping its real part,  potentially gives a debiased estimator for $f(X)$ that improves upon $f(W)$. Define $\hat f_{\text{pc}}(X)=\Re\{f(W-E(U)-i\{V-E(V)\})\}$ as this estimator involving a {\underline p}roper {\underline c}omplex random variable. 

A formal justification of the debiasing effect of $\hat f_{\text{pc}}(X)$ requires a comparison between the bias of this estimator (conditional on $X$), 
\begin{align}
    \label{eq:properCbias}
    \text{Bias}\{\hat f_{\text{pc}}(X)\} & = \sum_{\ell=3}^\infty \frac{f^{(\ell)}(X)}{\ell !} \Re\{E(C^\ell)\},
\end{align}
and the bias of the naive estimator $f(W)=f(X+U)$, elaborated as $f(X)+\sum_{\ell=1}^\infty \{f^{(\ell)}(X)/\ell!\} U^\ell$, is equal to 
\begin{align}
    \label{eq:naivebias}
    \text{Bias}\{f(W)\} & = \sum_{\ell=1}^\infty \frac{f^{(\ell)}(X)}{\ell !} E(U^\ell).
\end{align}
Even with $E(U)=0$ so that the infinite sum in \eqref{eq:naivebias} can start at $\ell=2$, $\hat f_{\text{pc}}(X)$ may still achieve some bias reduction because the infinite sum in \eqref{eq:properCbias} begins with $\ell=3$, and, when the $\ell$th moment is nonzero for $\ell>2$, $\Re\{E(C^\ell)\}$ is comparable with $E(U^\ell)$ since the highest-order moment of $U$ that $\Re\{E(C^\ell)\}$ linearly depends on is $E(U^\ell)$. 

Similar to the Monte Carlo corrected score estimator $\hat{f}_{\text{mc}}(X)$, one may use a Monte Carlo average,  $\hat{f}_{\text{pc}}(X)=\sum_{m=1}^M \Re\{f(W-E(U)+i\{V_m-E(V)\})\}/M$, to estimate $f(X)$ provided that one can generate $\{V_m\}_{m=1}^M$ from the measurement error distribution. On the other hand, \eqref{eq:properCbias} indicates that the propriety requirement on $C$ can be relaxed because one only needs $\Re\{E(C^\ell)\}=0$ instead of $E(C^\ell)=0$, for $\ell=1, 2$, when a real-valued estimator for $f(X)$ is desired. In fact, setting $C=U-E(U)+i\sigma_u $ suffices even though it is not a proper complex random variable. This gives another estimator for $f(X)$ that depends on a {\underline c}omplex {\underline c}onstant besides $W$, 
 \begin{equation}
 \label{eq:fcc}
 \hat{f}_{\text{cc}}(X)=\Re\{f(W-E(U)+i\sigma_u)\},
 \end{equation}
 with the same bias as $\hat{f}_{\text{pc}}(X)$ but freeing one from generating $\{V_m\}_{m=1}^M$, which in turn lowers the variability in estimation. To estimate $f(X)$ via \eqref{eq:fcc} only requires the first two moments of $U$ instead of its distribution as for $\hat{f}_{\text{mc}}(X)$ and $\hat{f}_{\text{pc}}(X)$. Certainly, in an application, one typically has to estimate these moments based on replicate measures of $X$ or some form of external or validation data. Even in these practical settings, estimating the first two moments is much easier and subject to less uncertainty than estimating the (unspecified) distribution of $U$.

 \subsection{Correcting for non-Gaussian measurement error using quaternions}
 Continuing on the course pursued in Section~\ref{sec:fcc}, one may wonder if it is possible to construct $C$ in \eqref{eq:taylor} with vanishing $\Re\{E(C^\ell)\}$ also for $\ell$ above 2 to improve over $\hat{f}_{\text{cc}}(X)$. As an extension of complex numbers that we utilize to achieve $\Re\{E(C)\}=\Re\{E(C^2)\}=0$, a natural place to look for possible solutions is the quaternion number system \citep{quaternion}. A quaternion $q$ is represented as $q = a + b i + c j + d k$, where $a,b,c,d \in\mathbb{R}$, and $i$, $j$, $k$ are the imaginary units that satisfy $i^2=j^2=k^2=- 1$, $ij=-j i=k$, $jk=-k j = i$, and $ k i =-i k =j$. These multiplication rules for the imaginary units imply that quaternion multiplication is noncommutative. 
 
 Consider letting $C=U+a+bi + cj +dk$ in \eqref{eq:taylor}. We now seek for $a, b, c, d\in \mathbb{R}$ such that $\Re\{E(C^\ell)\}=0$, for $\ell=1, 2, 3, 4$, if possible. Straightforward quaternion arithmetic leads to 
 \begin{align}
     \Re\{E(C)\} & = E(U)+a, \label{eq:finda}\\
     \Re\{E(C^2)\} & = E\{(U+a)^2\}-(b^2+c^2+d^2), \label{eq:ReC2} \\
     \Re\{E(C^3)\} & =E\{(U+a)^3\}-3(b^2+c^2+d^2)E(U+a), \label{eq:ReC3}\\
     \Re\{E(C^4)\} & =E\{(U+a)^4\}-6(b^2+c^2+d^2)E\{(U+a)^2\}+(b^2+c^2+d^2)^2. \label{eq:ReC4}
 \end{align}
Clearly, $a=-E(U)$ makes $\Re\{E(C)\}=0$ by \eqref{eq:finda}. Setting $a=-E(U)$ in \eqref{eq:ReC2} and \eqref{eq:ReC3} reveals that $\Re\{E(C^2)\}=0$ amounts to $b^2+c^2+d^2=\sigma_u^2$ and $\Re\{E(C^3)\}=0$ is equivalent to $\mu_3=0$; lastly, \eqref{eq:finda} and \eqref{eq:ReC2} together indicate that \eqref{eq:ReC4} reduces to zero when the kurtosis $\mu_4/\sigma_u^4=5$, where $\mu_\ell$ denotes the $\ell$th central moment of $U$, for $\ell\in \mathbb{N}$. Therefore, if one sets $a=-E(U)$ to make $\Re\{E(C)\}=0$, and chooses $(b, c, d)$ to achieve $\Re\{E(C^2)\}=0$, then $\Re\{E(C^3)\}\ne 0$ whenever $U$ is asymmetric around its mean (thus $\mu_3 \ne 0$), and $\Re\{E(C^4)\}\ne 0$ if its kurtosis is not equal to 5. Consequently, further bias reduction than what is already achieved by $\hat f_{\text{cc}}(X)$ is generally unattainable by using a quaternion $C$. If $U$ is symmetric around its mean, setting $C=U-E(U)-i\sigma_u$ as in $\hat f_{\text{cc}}(X)$ already yields $\Re\{E(C^3)\}=\mu_3=0$. 

The reason for our failed attempt of using a quaternion to further reduce the bias of $\hat f_{\text{cc}}(X)$ is that, for $C=U-E(U)+bi+cj+dk$, $\Re\{E(C^\ell)\}$ depends on $(b, c, d)$ only via $b^2+c^2+d^2$, for $\ell\in \mathbb{N}\setminus\{1\}$. Hence, after one achieves $\Re\{E(C^2)\}=0$ via a certain choice of $(b, c, d)$, setting $\Re\{E(C^\ell)\}=0$ for an $\ell$ above 2 only reveals the moment condition(s) relating to $U$ under which the higher-order real moment of $C$ vanishes. This very reason motivates us to consider a different kind of hypercomplex number known as tessarines \citep{tessarine}.

\subsection{Correcting for non-Gaussian measurement error using tessarines}
\label{sec:ftc}
A tessarine $\mathcal{S}$ is formulated as
$\mathcal{S}=a + b i+c j+d k$, where $a,b,c,d \in\mathbb{R},$ and  $i$, $j$, $k$ are the imaginary units that satisfy $i^2=-j^2=k^2=-1$, $i j= j i=k$, $j k=k j =i$, and $k i = i  k =-j$. To distinguish the imaginary units following these multiplication rules that are slightly different from those for the quaternion imaginary units, we use $\hat{i}$, $\hat{j}$, and $\hat{k}$ to denote the imaginary units in tessarines henceforth. These rules suggest that tessarine multiplication is commutative and that tessarines have zero divisors, that is, the product of two non-zero tessarines can equal zero. 

For a tessarine random variable $C=U+a+b\hat i+c \hat j+d\hat k$, the first real-valued moment is the same as \eqref{eq:finda}, but, unlike  \eqref{eq:ReC2}--\eqref{eq:ReC4}, we now have, following simple tessarine arithmetic,  
\begin{align}
     \Re\{E(C^2)\}= &\ E\{(U+a)^2\}-(b^2-c^2+d^2), \label{eq:ReC2t} \\
     \Re\{E(C^3)\} = &\ E\{(U+a)^3\}-3(b^2-c^2+d^2)E(U+a)-6bcd, \label{eq:ReC3t}\\
     \Re\{E(C^4)\} = &\ E\{(U+a)^4\}-6(b^2-c^2+d^2)E\{(U+a)^2\}-24 bcd E(U+a) \nonumber \\
     &\ +(b^2-c^2+d^2)^2-4\{b^2(c^2-d^2)+c^2d^2\}. \label{eq:ReC4t}
\end{align}
Like before, setting $a=-E(U)$ makes $\Re\{E(C)\}=0$. Plugging in \eqref{eq:ReC2t}--\eqref{eq:ReC4t} this choice of $a$ and setting them to zeros reveal that  
$(b, c, d)$ solves the following system of equations,
\begin{equation}
\label{eq:findbcd}
\begin{aligned}
    \sigma_u^2-b^2+c^2-d^2 & =0, \\
   \mu_3- 6bcd & =0 , \\
   \mu_4-5 \sigma_u^4-4\{ b^2(c^2-d^2)+c^2d^2\} & =0.
\end{aligned}
\end{equation}
And the estimator that involves the corresponding {\underline t}essarine {\underline c}onstant is \begin{equation}
    \label{eq:ftc}
     \hat{f}_{\text{tc}}(X)=\Re\{f(W-E(U)+b \hat i+c \hat j+d \hat k)\}.
 \end{equation}
Setting $b=\sigma_u$ and $c=d=0$ in \eqref{eq:ftc} makes $\hat f_{\text{tc}}(X)$ reduce to $\hat f_{\text{cc}}(X)$ in \eqref{eq:fcc}. If $f^{(3)}(t)=0$, for all $t\in \mathbb{R}$, then $\hat f_{\text{tc}}(X)$ has the same bias as that of $\hat f_{\text{cc}}(X)$ since all terms in the Taylor series with $\ell\ge 3$ in \eqref{eq:properCbias} vanish despite the values of $\Re\{E(C^\ell)\}$. As long as $f^{(3)}(t) \ne 0$ for some $t\in \mathbb{R}$,  $\hat f_{\text{tc}}(X)$ can potentially improve over $\hat{f}_{\text{cc}}(X)$. 
 Computing $\hat f_{\text{tc}}(X)$ requires evaluating $f(\cdot)$ at a tessarine and finding $(b, c, d)$ that satisfies \eqref{eq:findbcd}, which we elaborate next.

 \section{Implementation}
 \label{sec:implement}
 \subsection{Computing tessarine functions}\label{sec:bicomplex} 
Deriving $\Re\{E(C^\ell)\}$ when $C$ is a tessarine in Section~\ref{sec:ftc} is straightforward because computing the polynomial of a tessarine only involves the multiplication rules of tessarine imaginary units and simple tessarine arithmetic. It is however less obvious how to compute, for instance, $\exp(C)$. As being done for complex numbers and quaternions \citep{jacobson2012basic, zhang1997quaternions}, it is convenient to use a $2\times 2$ matrix with complex-valued entries to represent a tessarine so that every operation that can be done on tessarines can translate to an operation on matrices. This alternative representation of a tessarine is based on the matrix representation of the tessarine basis elements, 1, $\hat i$, $\hat j$, $\hat k$, given by $\mathbbm{1}=\bI_2$, $\mathbb{I}=i\bI_2$, $\mathbb{J}=\bJ_2$, and $\mathbb{K}=i\bJ_2$, respectively, where $\bI_2$ is the  $2\times 2$ identify matrix, $\bJ_2$ results from swapping the columns of $\bI_2$, and $i=(-1)^{1/2}$. One can easily verify that the basis elements in the matrix form, $\mathbbm{1}$, $\mathbb{I}$, $\mathbb{J}$, and $\mathbb{K}$, satisfy the multiplication rules for the original basis elements, that is, $\mathbb{I}^2=-\mathbb{J}^2=\mathbb{K}^2=-\mathbbm{1}$, $\mathbb{I}\mathbb{J} =\mathbb{J}\mathbb{I}=\mathbb{K}$, $\mathbb{J}\mathbb{K}=\mathbb{K}\mathbb{J} =\mathbb{I}$, and $\mathbb{K} \mathbb{I} =\mathbb{I}\mathbb{K} =-\mathbb{J}$. Despite the noncommutativity of matrix multiplication in general, the commutativity of tessarine multiplication is preserved under the matrix representation of tessarines. In the sequel, we view $C=a+b\hat i +c \hat j +d \hat k$ equivalent to the $2 \times 2$ complex matrix $C=a \mathbbm{1}+ b \mathbb{I}+c\mathbb{J}+d\mathbb{K}$, with $\Re(C)=a$ appearing in (the real parts of) the diagonal entries. With this matrix representation of $C$, the reciprocal of $C$ is equivalent to the inverse of the corresponding matrix, and the product of tessarines is reflected as the product of matrices. The reason for the presence of zero divisors in the tessarine system also becomes clear since zero divisors for matrices are a common occurrence. 

As an application of tessarine multiplication, the following lemma relates the tessarine power function to matrix power.
 \begin{lemma}
 \label{lem:power}
     For a tessarine 
     $$C=a \mathbbm{1}+ b \mathbb{I}+c\mathbb{J}+d\mathbb{K}=(a+bi)\bI_2+(c+di)\bJ_2=\begin{bmatrix} a+b i && c+ di \\ c+ di && a+bi \end{bmatrix},$$ 
     the power function evaluated at $C$ results in another tessarine $C^\ell$ equal to, for $\ell\in \mathbb{N}$,
     \begin{align*}
         &\ \frac{1}{2}\left(\left[\{a-c+i(b-d)\}^\ell+\{a+c+i(b+d)\}^\ell\right]\bI_2+ \right. \\
         &\ \left.\left[\{a+c+i(b+d)\}^\ell-\{a-c+i(b-d)\}^\ell \right] \bJ_2\right);
     \end{align*}  
     and consequently, 
      \begin{equation}
      \label{eq:ReCpower}
     \begin{aligned}
     \Re(C^\ell) =&\ \frac{1}{2}\left[\left\{(a-c)^{2}+(b-d)^{2}\right\}^{\ell/2}\cos(\ell\times \text{arg}\{a-c+i(b-d)\})\right.\\
     &\ \left.+\left\{(a+c)^{2}+(b+d)^{2}\right\}^{\ell/2}\cos(\ell \times\text{arg}\{a+c+i(b+d)\})\right],
     \end{aligned}
     \end{equation}
     where $\text{arg}(t)$ is the  argument of a complex number $t$.
 \end{lemma}
The proof of Lemma~\ref{lem:power} is provided in Appendix~\ref{app:provelem2}, starting with matrix diagonalization for $C$. Using Lemma~\ref{lem:power}, we prove Theorem~\ref{thm:tesexp} in Appendix~\ref{app:provethm3.1}, where we use the matrix exponential to derive the tessarine exponential. 
 \begin{theorem}\label{thm:tesexp}
     For a tessarine $C=a+b\hat i+c \hat j +d \hat k$,  
    $$\exp(C) 
     =\exp(a+bi)\{\cosh(c+di)\bI_2+\sinh(c+di)\bJ_2\},$$
     where $\sinh(t)=(e^t-e^{-t})/2$ and  $\cosh(t)=(e^t+e^{-t})/2$
     for a complex $t$.  
     Thus
     $\Re\{\exp(C)\}=\exp(a)\{\cos(b)\cos(d)\cosh(c)-\sin(b)\sin(d)\sinh(c)\}$.
 \end{theorem}
 
Using Theorem~\ref{thm:tesexp}, one can evaluate any function at a tessarine that can be expressed via exponential functions, such as trigonometric functions. If $f(\cdot)$ is not any of the polynomials, exponential, and trigonometric functions, one may resort to the Taylor series to evaluate the estimator in \eqref{eq:ftc} under the entirety assumption for $f(\cdot)$, that is, $\hat f_{\text{tc}}(X)=\sum_{\ell=0}^\infty \{f^{(\ell)}(W-E(U))/\ell !\}\Re\{(b\hat i+c \hat j +d\hat k)^\ell\}$, and one can use \eqref{eq:ReCpower} with $a=0$ to compute the summand.

\subsection{Finding the tessarine}
\label{sec:algo}
Following the resultant procedure  \citep{resultants}, one can show that \eqref{eq:findbcd} is equivalent to 
\begin{align}
   \mu_3^2+9(5\sigma_u^4-\mu_4)d^2-36\sigma_u^2 d^4+36d^6 & =0, \label{eq:tofindd}\\
\mu_3^{2}+36d^2(d^2-\sigma_u^2)c^2-36d^2c^4 & =0, \label{eq:tofindc} \\
        b^2-c^2+d^2-\sigma_u^2 & =0 \label{eq:tofindb}.
\end{align}
One can find $d$, $c$, and $b$ sequentially, first solving \eqref{eq:tofindd} for $d$, then solving \eqref{eq:tofindc} for $c$,  and lastly solving \eqref{eq:tofindb} for $b$, each step using a simple root-finding algorithm. This provides a simple algorithm for finding a real-valued solution to \eqref{eq:findbcd} when real solutions exist.  

If there is no real solution to \eqref{eq:findbcd}, then there is no tessarine of the form $C=U-E(U)+b\hat i+c\hat j +d\hat k$ satisfying $\Re\{E(C^2)\}=\Re\{E(C^3)\}=\Re\{E(C^4)\}=0$. To reduce the bias in \eqref{eq:properCbias}, we now aim to find $(b, c, d)\in \mathbb{R}^3$ that minimizes these real moments in combination by minimizing the following objective function,
\begin{equation}\label{obj}
    Q(b,c,d)=\bigg\{\frac{f_{1}(b,c,d)}{\sigma_u^2}\bigg\}^2+\bigg\{\frac{f_{2}(b,c,d)}{3\mu_3}\bigg\}^2+\bigg\{\frac{f_{3}(b,c,d)}{12\mu_4}\bigg\}^2,
\end{equation}
where $f_1(b, c, d)$ is the difference between the two sides of \eqref{eq:ReC2t} with $a$ set at $-E(U)$, $f_2(b, c, d)$ and $f_3(b, c, d)$ are similarly defined based on \eqref{eq:ReC3t} and \eqref{eq:ReC4t}, respectively, and the weights in \eqref{obj} (proportional to $\ell !$ for $\ell=2, 3, 4$) are motivated by the Taylor series in \eqref{eq:properCbias}. One may employ the gradient descent algorithm to find a minimizer of \eqref{obj}, using multiple starting points to avoid being trapped at a saddle point or local minimum. 

\section{Applying to M-estimation in regression analysis}
\label{sec:Mest}
\subsection{Overview}
Consider the generic problem of M-estimation based on data from $n$ independent experimental units, $\{(Y_r, X_r)\}_{r=1}^n$, that one uses to fit a regression model with $\Theta$ as the parameter(s) of interest. A consistent M-estimator for $\Theta$ solves the estimating equation, 
$\sum_{r=1}^{n}{\Psi}(Y_r,X_r,\Theta)=0$, where $\Psi(\cdot)$ is a score function satisfying $E\{{\Psi}(Y,X,\Theta^*)\}=0$, in which $\Theta^*$ is the true value of $\Theta$ \citep{stefanski2002calculus}, and we suppress the index $r$ in $Y$ and $X$ (and other random quantities later) when we refer to the data of an experimental unit generically. In the presence of measurement error when $\{W_r\}_{r=1}^n$ are observed instead of $\{X_r\}_{r=1}^n$, the corrected score method \citep{nakamura} suggests using a different score function that is an unbiased estimator of ${\Psi}({Y},{X},{\Theta})$ given $(Y, X)$ in the estimating equation. Using the notations in earlier sections, $f(X)={\Psi}({Y},{X},{\Theta})$, and its unbiased estimator is referred to as a corrected score. 

Finding a corrected score usually has to be done on a case-by-case basis, which is a daunting task in most regression settings, and it may not be possible to find one explicitly. Without assuming Gaussian measurement error in \eqref{eq:addme}, we propose to estimate ${\Psi}({Y},{X},{\Theta})$ via  
\begin{equation}
\label{eq:gencs}
\tilde{{\Psi}}({Y},{W},{{\Theta}})=
   \Re\{{\Psi}(Y, \, W-E(U)+\Im, \, {{\Theta}})\},
\end{equation}
where $\Im$ is the imaginary portion of a (hyper)complex number considered in Section~\ref{sec:method}. More specifically, setting $\Im=i \sigma_u$ as in \eqref{eq:fcc} gives a score as an application of $\hat f_{\text{cc}}(X)$, and having $\Im=b\hat{i}+c\hat{j}+d\hat{k}$ leads to another score in the spirit of $\hat f_{\text{tc}}(X)$, with $b,c,d$ obtained as described in Section~\ref{sec:algo} based on assumed or estimated first four moments of $U$. The goal here is to abort the often impossible mission of analytically finding an unbiased estimator for the true-data score $\Psi(Y, X, \Theta)$, and instead to find a score that improves over the naive observed-data score $\Psi(Y, W, \Theta)$. We call a score function in the form of \eqref{eq:gencs} a debiased score in contrast with a corrected score. We zoom in on four regression models to illustrate this new score next. 

\subsection{Linear Regression}
\label{sec:linearreg}
Consider a linear regression model with error-in-covariates,
\begin{equation*}
    Y_r=\beta_0+X_r^\top \beta+\epsilon_r,\qquad W_r=X_r+U_r, \quad (r=1, \ldots, n),
\end{equation*}
where $\Theta=(\beta_0, \beta^\top)^\top\in \mathbb{R}^{m+1}$ contains the parameters of interest, $\epsilon_r$ is a mean-zero error, and the additive measurement error model generalizes \eqref{eq:addme} by allowing a multivariate covariate supported on $\mathbb{R}^m$, and also the $m$-dimensional measurement error $U_r$ with the variance-covariance matrix denoted by $\Omega$. In the absence of measurement error, the least square estimator of $\Theta$ is the solution to the estimating equation $\sum_{r=1}^n (Y_r-\beta_0-X_r^\top \beta)(1, X_r^\top)^\top=0_{m+1}$, where $0_{m+1}$ is the $(m+1)\times 1$ vector of zeros. That is, the true-data score is $\Psi(Y, X, \Theta)=(Y-\beta_0-X^\top \beta)(1, X^\top)^\top$. 

In the presence of measurement error, the naive score, $\Psi(Y, W, \Theta)=(Y-\beta_0-W^\top \beta)(1, W^\top)^\top$, is typically a biased estimator of $\Psi(Y, X, \Theta)$ given $(Y, X)$, with the bias provided by the multivariate version of the Taylor series in \eqref{eq:naivebias}. Because $f(X)=\Psi(Y, X, \Theta)$ is a quadratic function of $X$ here, terms of order above two (i.e., $\ell>2$) in the Taylor series reduce to zero, $\hat f_{\text{tc}}(X)$ does not achieve further bias reduction compared to $\hat f_{\text{cc}}(X)$. The application of $\hat f_{\text{cc}}(X)$ in the presence of multivariate measurement error is equivalent to setting $\Im=iB$ in \eqref{eq:gencs}, where $B=(b_1, \ldots, b_m)^\top \in \mathbb{R}^m$ is chosen to reduce, or ideally, eliminate the bias of $\Psi(Y, W, \Theta)$. One can generalize the arguments in Section~\ref{sec:method} to adapt to an $m$-dimensional complex random variable, $C=U-E(U)+iB$, and find $B$ such that $\Re\{E(C^2)\}$ vanishes, where $C^2=C_1^{\ell_1}\cdots C_m^{\ell_m}$, in which $C_1, \ldots, C_m$ are entries of $C$, and $\ell_1, \ldots, \ell_m \in \{0, 1, 2\}$ such that $\ell_1+\ldots+\ell_m=2$. To gain more insight on the debiased score $\tilde \Psi(Y, W, \Theta)$ in \eqref{eq:gencs} under linear regression, we take a different route to find $B$ via elaborating \eqref{eq:gencs} next. 

Define $\mathscr{T}=W-E(U)+\Im$. With the least square score function in \eqref{eq:gencs}, we have 
\begin{align}
\tilde \Psi(Y, W, \Theta) & = \Re\left\{
(Y-\beta_0-\mathscr{T}^\top \beta) 
\begin{bmatrix}
    1 \\ \mathscr{T} 
\end{bmatrix}
\right\} \nonumber \\
 & = \left[Y-\beta_0 -\{W-E(U)\}^\top\beta \right]
 \begin{bmatrix}
     1 \\
     W-E(U)
 \end{bmatrix}-
 \begin{bmatrix}
     0 \\
     \Re(\Im\Im^\top) \beta
 \end{bmatrix}. \label{eq:linregsc}
\end{align}
Taking the expectation of \eqref{eq:linregsc} conditional on $(Y, X)$ gives 
\begin{align*}
    E\{\tilde \Psi(Y, W, \Theta)|Y, X\} & = 
\begin{bmatrix}
     Y-\beta_0-X^\top \beta \\
     (Y-\beta_0 -X^\top \beta)X -\Omega\beta -\Re(\Im\Im^\top)\beta
\end{bmatrix},
\end{align*}
which is equal to the least square score $\Psi(Y, X, \Theta)$ 
when $\Im=iB$, with $B$ satisfying $BB^\top=\Omega$. This gives a debiased score, $\tilde \Psi(Y, W, \Theta)=[Y-\beta_0\{W-E(U)\}^\top \beta](1, \{W-E(U)\}^\top)^\top+(0, \beta^\top \Omega)^\top$, which  
 coincides with the corrected score in linear regression with error-in-covariates, and includes $\hat f_{\text{cc}}(X)$ in \eqref{eq:fcc} as a special case when $m=1$. 

\subsection{Polynomial regression}
\label{sec:polyreg}
Consider an $m$th order polynomial regression model with error-in-covariate, 
$$Y_r=\beta_{0}+\sum_{q=1}^{m}\beta_{q}X_r^q+\epsilon_r,\qquad W_r=X_r+U_r,\qquad (r=1,\ldots,n),$$
where $\Theta=(\beta_0, \beta_1, \ldots, \beta_m)^\top\in \mathbb{R}^{m+1}$ is the parameter vector of interest, $\epsilon_r$ is a mean-zero error, and the measurement error model is the same as \eqref{eq:addme} with a univariate nondifferential measurement error. The least square estimator of $\Theta$ in the absence of measurement error solves the estimating equation $\sum_{r=1}^n (Y_r-\beta_0-\sum_{q=1}^m \beta_q X_r^q)(1, X_r, X_r^2, \ldots, X_r^m)^\top=0_{m+1}$.  That is, the true-data score is $\Psi(Y, X, \Theta)=Y\tilde X-\tilde X \tilde X^\top \Theta$, where $\tilde X=(1, X, X^2, \ldots, X^m)^\top$. Similarly, let $\tilde W=(1, W, W^2, \ldots, W^m)^\top$.

Because the score function $f(X)=\Psi(Y, X, \Theta)$ here is a polynomial function of order $2m$ of $X$, $f^{(3)}(t)$ is not always zero when $m\ge 2$. According to \eqref{eq:naivebias}, the bias of the naive score, $\Psi(Y, W, \Theta)=Y\tilde W-\tilde W \tilde W^\top \Theta$, depends on higher-order moments of $U$, $\mu_\ell$, for $\ell \ge 3$. This is the scenario where $\hat f_{\text{tc}}(X)$ can improve over $\hat f_{\text{cc}}(X)$. We thus consider a debiased score in \eqref{eq:gencs} involving tessarines, that is, 
\begin{equation}
\tilde \Psi(Y, W, \Theta)=\Re(Y\tilde{\mathscr{T}}-\tilde{\mathscr{T}} \tilde{\mathscr{T}}^\top \Theta)= Y\Re(\tilde{\mathscr{T}})-\Re(\tilde{\mathscr{T}} \tilde{\mathscr{T}}^\top) \Theta, \label{eq:debscpoly}    
\end{equation}
where $\tilde{\mathscr{T}}=(1, \mathscr{T}, \mathscr{T}^2, \ldots, \mathscr{T}^m)^\top$, in which $\mathscr{T}=W-E(U)+b\hat i+c\hat j+d\hat k$. After one finds $(b, c, d)$ following the strategies described in Section~\ref{sec:algo}, one can use \eqref{eq:ReCpower} to compute $\Re(\mathscr{T}^\ell)$, for $\ell=2, 3, \ldots, 2m$, and then evaluates \eqref{eq:debscpoly}.

Let us take the quadratic regression as an example for further elaboration. With $m=2$, we have $\Re(\tilde {\mathscr{T}})$ equal to 
\begin{equation*}
    \Re\left( 
    \begin{bmatrix}
        1 \\
        \mathscr{T}\\
        \mathscr{T}^2
    \end{bmatrix}
    \right) = 
    \begin{bmatrix}
        1\\
        W-E(U) \\
        \{W-E(U)\}^2-(b^2-c^2+d^2)
    \end{bmatrix}
    = 
    \begin{bmatrix}
        1\\
        W-E(U) \\
        \{W-E(U)\}^2-\sigma_u^2
    \end{bmatrix},     
\end{equation*}
where we conclude the third entry of $\Re(\tilde {\mathscr{T}})$ by \eqref{eq:tofindb}. The $[s, t]$ entry of $\Re(\tilde{\mathscr{T}}\tilde{\mathscr{T}}^\top)$ is $\Re( \mathscr{T}^{s+t-2})$, for $s, t\in \{1, 2, 3\}$. That is, entries of $\Re(\tilde{\mathscr{T}}\tilde{\mathscr{T}}^\top)$ include those in $\Re(\tilde {\mathscr{T}})$ and 
\begin{align*}
    \Re(\mathscr{T}^3)& = \{W-E(U)\}^3-3 (b^2-c^2+d^2)\{W-E(U)\}-6bcd \\
    & = \{W-E(U)\}^3-3\sigma_u^2\{W-E(U)\}-\mu_3,\\
    \Re(\mathscr{T}^4)& = \{W-E(U)\}^4-6 (b^2-c^2+d^2)\{W-E(U)\}^2-24bcd\{W-E(U)\}\\
    &\ \quad -4\{b^2(c^2-d^2)+c^2d^2\}+(b^2-c^2+d^2)^2 \\
    & = \{W-E(U)\}^4-6\sigma_u^2\{W-E(U)\}^2-4\mu_3\{W-E(U)\}-\mu_4+6\sigma_u^4,
\end{align*}
where constraints on $(b, c, d)$ suggested in \eqref{eq:findbcd} are used to simplify both results. Plugging in  \eqref{eq:debscpoly}  the resultant $\Re(\tilde {\mathscr{T}})$ and $\Re(\tilde{\mathscr{T}}\tilde{\mathscr{T}}^\top)$ yields the debiased score $\tilde \Psi(Y, W, \Theta)$ that is identical to the corrected score provided in \cite{polynomial} for a quadratic regression model. In \cite{polynomial}, the authors derived unbiased estimators of $Y\tilde X$ and $\tilde X \tilde X^\top$ to construct an unbiased estimator of $\Psi(Y, X, \Theta)=Y\tilde X-\tilde X \tilde X^\top \Theta$, for $m\ge 2$, without assuming $\epsilon$ and $U$ independent or assuming Gaussian measurement error. The price they pay for this generality is the assumption that $\{\mu_\ell\}_{\ell=1}^{2m}$ and $\{E(U^\ell \epsilon)\}_{\ell=1}^m$ are all known. The tessarine-based debiased score $\hat f_{\text{tc}}(X)$ is a corrected score only when $m\le 2$, but can still reduce bias when $m>2$ in the naive score $f(W)$, the Monte Carlo corrected score $\hat f_{\text{mc}}(X)$ in the presence of non-Gaussian measurement error, and the score involving a complex constant, $\hat f_{\text{cc}}(X)$, while only requiring knowledge of $\{\mu_\ell\}_{\ell=1}^4$.

\subsection{Poisson regression}
\label{sec:poisreg}
Consider a Poisson regression model  \citep{Poissonbook} with error-in-covariate, 
\begin{equation}
Y_r\mid X_r\sim \text{Poisson}(\exp(\beta_{0}+\beta_{1}X_r)), \qquad W_r=X_r+U_r, \qquad (r=1,\ldots,n),\label{eq:poisreg}
\end{equation}
where $\Theta=(\beta_0, \beta_1)^\top\in \mathbb{R}^2$ contains regression coefficients of interest. Had there been no measurement error, the maximum likelihood estimator of $\Theta$ is the solution to the estimating equation $\sum_{r=1}^n \{Y_r-\exp(\beta_0+\beta_1 X_r)\}(1, X_r)^\top=0_2$. Hence, the true-data score function is $\Psi(Y, X, \Theta)=\{Y-\exp(\beta_0+\beta_1 X)\}(1, X)^\top$. 

To reduce the bias of the naive score $\Psi(Y, W, \Theta)=\{Y-\exp(\beta_0+\beta_1 W)\}(1, W)^\top$ in the presence of non-Gaussian measurement error, we consider the debiased score in \eqref{eq:gencs} that involves tessarine $\mathscr{T}=W-E(U)+b\hat i+c\hat j+d\hat k$, that is,
\begin{align}
\tilde \Psi(Y, W, \Theta) & =\Re\left(\{Y-\exp(\beta_0+\beta_1 \mathscr{T})\}
\begin{bmatrix}
    1 \\ \mathscr{T}
\end{bmatrix}\right)  \nonumber \\
 & = 
\begin{bmatrix}
    Y-e^{\beta_0}\Re\{\exp(\beta_1 \mathscr{T})\} \\
    Y\{W-E(U)\}-e^{\beta_0}\Re\{\exp(\beta_1 \mathscr{T})\mathscr{T}\}
\end{bmatrix},
\label{eq:poisdebscore}
\end{align}
where, using Theorem~\ref{thm:tesexp}, we have
\begin{align*}
    &\ \Re\{\exp(\beta_1 \mathscr{T})\} \\ 
    = &\ \frac{1}{2}\exp(\beta_1\{W-E(U)\})\left\{e^{\beta_1c}\cos(\beta_1(b+d))+e^{-\beta_1c}\cos(\beta_1(b-d)) \right\},
\end{align*}
and, using the matrix representations of $\mathscr{T}$ (as in Lemma~\ref{lem:power}) and $\exp(\beta_1 \mathscr{T})$ (given in Theorem~\ref{thm:tesexp}) to carry out the tessarine multiplication $\exp(\beta_1 \mathscr{T}) \mathscr{T}$ as matrix multiplication, then extracting the real part of the first diagonal entry of the resultant  matrix gives  
\begin{align*}
    &\ \Re\{ \exp(\beta_1 \mathscr{T}) \mathscr{T}\} \\
    = &\ \frac{1}{2}\exp(\beta_1\{W-E(U)\})\big(e^{\beta_1c}[\{W-E(U)+c\}\cos(\beta_1(b+d))\\
    &\ -(b+d)\sin(\beta_1(b+d))]
    +e^{-\beta_1c}[\{W-E(U)-c\}\cos(\beta_1(b-d))\\
    &\ +(b-d)\sin(\beta_1(b-d)) ]\big).
\end{align*}
To express the debiased score more succinctly, we assume $E(U)=0$ and let $\xi_1=\cos(\beta_1(b+d))$, $\xi_2=\cos(\beta_1(b-d))$, 
$\eta_1=\sin(\beta_1(b+d))$, and $\eta_1=\sin(\beta_1(b-d))$, 
then 
\begin{equation}
\label{eq:poisdebias}
\begin{aligned}
&\ \tilde \Psi(Y, W, \Theta)\\
= &\ 
\begin{bmatrix}
    Y-\exp(\beta_0+\beta_1W) \left( \xi_1 e^{\beta_1 c}+\xi_2 e^{-\beta_1 c}\right)/2 \\\\
    YW-\exp(\beta_0+\beta_1W)\left[\{\xi_1(W+c)-\eta_1(b+d)  \}e^{\beta_1c}\right.\\
   \left. +\{\xi_2(W-c)+\eta_2(b-d)  \}e^{-\beta_1c}\right]/2
\end{bmatrix}.
\end{aligned}
\end{equation}
If $U$ is symmetric around zero with the kurtosis greater than 5, then one can show that a solution to \eqref{eq:findbcd} is $(b, c, d)=(b^*, c^*, 0)$, with $b^*=\{\sigma_u^2+(\mu_4-4\sigma_u^4)^{1/2}\}^{1/2}/\sqrt{2}$ and $c^*=(b^{*2}-\sigma_u^2)^{1/2}$. Using this choice of $(b, c, d)$ in \eqref{eq:poisdebias} yields a debiased score that partially accounts for symmetric measurement error that has a heavier tail than a Gaussian distribution, 
\begin{equation*}
\begin{aligned}
&\ \tilde \Psi(Y, W, \Theta)\\
= &\ 
\begin{bmatrix}
    Y-\exp(\beta_0+\beta_1W)\cos(\beta_1 b^*)\cosh(\beta_1c^*)\\\\
    YW-\exp(\beta_0+\beta_1W)\left[\cos(\beta_1 b^*)\cosh(\beta_1c^*)W\right.+ \\
    \left.\{c^*\cos(\beta_1b^*)-b^* \sin(\beta_1 b^*)\}\sinh(\beta_1c^*)\right]
    \end{bmatrix}.
\end{aligned}
\end{equation*}
This new score is obviously very different from the following \underline{c}orrected \underline{s}core derived under the assumption of mean-zero Gaussian measurement error for Poisson regression \citep[][Section 7.4.3]{carroll2006measurement}, 
\begin{align}
    \Psi_{\text{cs}}(Y, W, \Theta) & = 
    \begin{bmatrix}
        Y-\exp(\beta_0+\beta_1 W-\beta_1^2\sigma_u^2/2) \\
        YW-\exp(\beta_0+\beta_1 W-\beta_1^2\sigma_u^2/2)(W-\beta_1 \sigma_u^2)
    \end{bmatrix}.\label{eq:poiscs}
\end{align}

\subsection{Logistic regression}
Having considered regression models for continuous responses and response as a count in Sections~\ref{sec:linearreg}--\ref{sec:poisreg}, we shift our focus to a logistic regression model for a binary response with error-in-covariate in this subsection,
\begin{equation}
    Y_r\mid X_r\sim\text{Bernoulli}(\mathcal{G}(\beta_{0}+\beta_{1}X_r)),\qquad W_r=X_r+U_r, \qquad (r=1, \ldots, n),
    \label{eq:logitreg}
\end{equation}
where $\Theta=(\beta_0, \beta_1)^\top\in \mathbb{R}^2$ is the parameter of interest, and $\mathcal{G}(t)=1/\{1+\exp(-t)\}$ is the expit function. Based on error-free data, the maximum likelihood estimator of $\Theta$ solves the estimating equation $\sum_{r=1}^n\{Y_r-\mathcal{G}(\beta_0+\beta_1 X_r)\}(1, X_r)^\top=0_2$. 

Unlike the true-data scores considered in earlier subsections, here we have $f(X)=\Psi(Y, X, \Theta)=\{Y-\mathcal{G}(\beta_0+\beta_1 X)\}(1, X)^\top$, which is not an entire function of $X$ because the expit function $\mathcal{G}(t)$ is not entire \citep{levin1996lectures}. In particular, $t=i\ell\pi$ is a singularity point of $\mathcal{G}(t)$ when $\ell$ is an odd integer. As a result, one cannot use the  Taylor series as those in \eqref{eq:taylor}--\eqref{eq:naivebias} to justify or inspect the debiasing effect of the score in \eqref{eq:gencs} when comparing with the naive score $f(W)=\Psi(Y, W, \Theta)=\{Y-\mathcal{G}(\beta_0+\beta_1 W)\}(1, W)^\top$. This complication due to the nonentire $f(X)$ presents the same challenge in applying the Monte Carlo corrected score $\hat f_{\text{mc}}(X)$ in logistic regression. For simpler notations, suppose a random sample of size $M=1$ from $N(0, 1)$ is used in constructing $\hat f_{\text{mc}}(X)$, then this score is  \citep[][Secton 8.2]{StefNovick2002} 
\begin{equation}
    \Psi_{\text{mc}}(Y, W, \Theta)=
    \begin{bmatrix}
        Y-\displaystyle{\frac{\exp(\beta_0+\beta_1 W)+\cos(\beta_1\sigma_u Z)}{\exp(-\beta_0-\beta_1 W)+\exp(\beta_0+\beta_1 W) +2\cos(\beta_1 \sigma_uZ)}} \\\\
        YW-\displaystyle{\frac{\{\exp(\beta_0+\beta_1 W)+\cos(\beta_1\sigma_u Z)\}W-\sigma_uZ\sin(\beta_1\sigma_uZ)}{\exp(-\beta_0-\beta_1 W)+\exp(\beta_0+\beta_1 W) +2\cos(\beta_1 \sigma_uZ)}}
    \end{bmatrix}, 
    \label{eq:logitmccs}
\end{equation}
where $Z\sim N(0, 1)$. The authors argued that, even though $ \Psi_{\text{mc}}(Y, W, \Theta)$ is a biased estimator for $\Psi(Y, X, \Theta)$ (thus is not a corrected score), the bias is small ``for measurement error variances of the magnitudes commonly encountered in applications.`` They provided empirical evidence suggesting that estimates for $\Theta$ resulting from the Monte Carlo corrected score method are similar to those from the conditional score method. The \underline{c}on\underline{d}itional score  is an unbiased estimator of $\Psi(Y, X, \Theta)$ (and thus is a corrected score) when $U$ is Gaussian; more specifically, this score is \citep[][Section 7.2.2]{carroll2006measurement}
\begin{equation}
 \Psi_{\text{cd}}(Y,W, \Theta)=\{Y-\mathcal{G}(\beta_0+(W+Y\sigma_u^2\beta_1)\beta_1-\beta_1^2\sigma_u^2/2)\}(1, W+Y\sigma_u^2 \beta_1)^\top.   
 \label{eq:logistcds}
\end{equation}

Inspired by these existing works on logistic regression with covariates contaminated by Gaussian measurement error, we consider the debiased score in \eqref{eq:gencs} with $\mathscr{T}=W-E(U)+b\hat i+c\hat j+d\hat k$ substituting $X$ in $\Psi(Y, X, \Theta)$, leading to 
\begin{align}
    \tilde \Psi(Y, W, \Theta) & = 
    \begin{bmatrix}
        Y-\Re\{\mathcal{G}(\beta_0+\beta_1 \mathscr{T})\} \\
        Y\{W-E(U)\}-\Re\{\mathcal{G}(\beta_0+\beta_1 \mathscr{T})\mathscr{T}\}
    \end{bmatrix}, \label{eq:logisttc}
\end{align}
with the two real parts elaborated in Appendix~\ref{app:logistic}.  
We will assess the debiasing effect of the corresponding estimator for $\Theta$ in the presence of non-Gaussian measurement error in simulation experiments in Section~\ref{sec:simu}.

\subsection{Asymptotic bias analysis}
Following the inspection of the debiased score in \eqref{eq:gencs} applied to four regression models, we now study the corresponding debiased estimator in general. Suppose there exists a solution $(b, c, d)$ to $\eqref{eq:findbcd}$ so that $\Re\{E(C^\ell)\}=0$ for $\ell=1, 2, 3, 4$, with $C=U-E(U)+b\hat i+c\hat j+d\hat k$. Denote by $\hat \Theta_{\text{tc}}$ the debiased estimator involving tessarine random errors, $\{C_r=U_r-E(U)+b\hat i+c\hat j+d\hat k\}_{r=1}^n$, that  satisfies $n^{-1}\sum_{r=1}^n \Re \{\Psi(Y_r, X_r+C_r, \hat \Theta_{\text{tc}})\}=0$. Assuming $\Psi(y, x, \theta)$ infinitely differentiable with respect to $x$ for each $(y, \theta)$, then the estimating equation is equivalent to, following the Taylor expansion of $\Psi(\cdot, X_r+C_r, \cdot)$ around $X_r$, 
\begin{align}
     \frac{1}{n}\sum_{r=1}^n \left\{\Psi(Y_r, X_r, \hat \Theta_{\text{tc}})+\sum_{\ell=1}^\infty \Psi^{(\ell)}_2(Y_r, X_r, \hat \Theta_{\text{tc}}) \Re\left(C_r^\ell\right)\right\}& =0, \label{eq:debiasesteq}
\end{align}
 where $\Psi^{(\ell)}_2(Y_r, x, \hat \Theta_{\text{tc}})=(\partial^\ell/\partial x^\ell)\Psi(Y_r, x, \hat \Theta_{\text{tc}})$, with the subscript ``2'' signifying that it is the $\ell$-th order partial derivative of $\Psi(y, x, \theta)$ with respect to the second argument. A first-order Taylor expansion of \eqref{eq:debiasesteq} around the truth $\Theta^*$ gives 
 \begin{align}
     0 = &\ \frac{1}{n}\sum_{r=1}^n \left\{\Psi(Y_r, X_r, \Theta^*)+\sum_{\ell=1}^\infty \Psi^{(\ell)}_2(Y_r, X_r,  \Theta^*) \Re\left(C_r^\ell\right)\right\} + \nonumber \\
     &\ \frac{1}{n}\sum_{r=1}^n \left\{\Psi_3^{(1)}(Y_r, X_r, \Theta^*)+\sum_{\ell=1}^\infty \Psi^{(\ell,1)}_{2,3}(Y_r, X_r,  \Theta^*) \Re\left(C_r^\ell\right)\right\}(\hat \Theta_{\text{tc}}-\Theta^*)+\mathcal{R}^*, \label{eq:asymbiastc}
 \end{align}
where $\Psi_3^{(1)}(y, x, \Theta^*)$ is $(\partial/\partial \theta)\Psi(y, x, \theta)$ evaluated at $\theta=\Theta^*$, with the subscript ``3'' stressing that it is the partial derivative with respect to the third argument of $\Psi(y, x, \theta)$; similarly, $\Psi^{(\ell,1)}_{2,3}(y, X_r,  \Theta^*)$ is equal to $(\partial/\partial \theta)\{(\partial^\ell/\partial x^\ell) \Psi(y, x, \theta)\}$ evaluated at $(x, \theta)=(X_r, \Theta^*)$; lastly, $\mathcal{R}^*$ is the remainder term of the Taylor expansion that depends on functionals (as derivatives) of $\Psi(y, x, \theta)$ evaluated at $(Y_r, X_r)$'s and $\tilde \Theta$ that lies between $\hat \Theta_{\text{tc}}$ and $\Theta^*$. 

By \eqref{eq:asymbiastc}, one has, assuming the invertibility of the first term below,
\begin{align}
    &\ \hat \Theta_{\text{tc}}-\Theta^* \nonumber \\
= &\ -\left\{\frac{1}{n}\sum_{r=1}^n \Psi_3^{(1)}(Y_r, X_r, \Theta^*)+\sum_{\ell=1}^\infty \frac{1}{n}\sum_{r=1}^n \Psi^{(\ell,1)}_{2,3}(Y_r, X_r,  \Theta^*) \Re\left(C_r^\ell\right)\right\}^{-1} \nonumber \\
&\ \times \left\{\frac{1}{n}\sum_{r=1}^n \Psi(Y_r, X_r, \Theta^*)+\sum_{\ell=1}^\infty \frac{1}{n}\sum_{r=1}^n\Psi^{(\ell)}_2(Y_r, X_r,  \Theta^*) \Re\left(C_r^\ell\right)+\mathcal{R}^*\right\} \label{eq:diffave} \\
= &\ -\left[E\left\{ \Psi_3^{(1)}(Y, X, \Theta^*)\right\}+\sum_{\ell=5}^\infty E\left\{\Psi^{(\ell,1)}_{2,3}(Y, X,  \Theta^*)\right\} \Re\left\{E\left(C^\ell\right)\right\}+o_p(1)\right]^{-1} \nonumber \\
&\ \times\left[\sum_{\ell=5}^\infty E\left\{\Psi^{(\ell)}_2(Y, X,  \Theta^*)\right\} \Re\left\{E\left(C^\ell\right)\right\}+\mathcal{R}^*+o_p(1)\right], \label{eq:diffexp}
\end{align}
under regularity conditions imposed on the score function $\Psi(y,x, \theta)$ (e.g, see   \citet[Theorem 7.1,][]{boos2013essential}), such as $E\left\{ \Psi(Y, X, \Theta^*)\right\}=0$ and moment conditions required for the empirical means in  \eqref{eq:diffave} converging in probability to the corresponding expectations in \eqref{eq:diffexp} by the weak law of large numbers. According to \eqref{eq:diffexp}, the dominating bias of $\hat \Theta_{\text{tc}}$ mainly depends on the partial derivatives of $\Psi(y, x, \theta)$ with respect to $x$ of order above four. A sufficient condition for this dominating term to vanish, besides all the aforementioned conditions leading to \eqref{eq:diffexp}, is to have these high-order derivatives vanish. The practical implication of this is that the bias of $\hat \Theta_{\text{tc}}$ tends to be more negligible when $\Psi(y, x, \theta)$, as a function of $x$, can be better approximated by some fourth-order polynomial for each $(y, \theta)$.

 \section{Kernel density estimation}
 \label{sec:density}
After considering in Section~\ref{sec:Mest} four parametric regression models with error-prone covariates, we turn to a nonparametric problem of estimating the probability density function of $X$, $p_X(x)$. In the absence of measurement error, an estimator for $p_X(x)$ based on a random sample $\{X_r\}_{r=1}^n$  is $\hat p_X(x)=(1/n)\sum_{r=1}^n K_h(X_r-x)$, where $K_{h}(\cdot)=(1/h)K(\cdot/h)$, $K(\cdot)$ is a kernel and $h>0$ is the bandwidth \citep{rosenblatt, parzen}. 

In the notations established in Section~\ref{sec:method},  here we have $f(X)=K_h(X-x)$, with the naive counterpart $f(W)=K_h(W-x)$ appearing in the naive density estimator of $p_X(x)$ based on the error-contaminated data $\{W_r\}_{r=1}^n$, $\hat p_{\text{naive}}(x)=(1/n)\sum_{r=1}^n K_h(W_r-x)$. The proposed debiased estimator of $f(X)$ in \eqref{eq:ftc} is then $\hat f_{\text{tc}}(X)=\Re\{K_h(W-x-E(U)+b\hat i+c\hat j +d\hat k)\}$ for some $(b, c, d)$ obtained according to Section~\ref{sec:algo}. This yields a new estimator of $p_X(x)$ that can improve over $\hat p_{\text{naive}}(x)$ in the presence of non-Gaussian measurement error, 
\begin{equation}
    \label{eq:denest}
    \hat p_{\text{tc}}(x) = \frac{1}{n}\sum_{r=1}^n \Re\left\{K_h(W_r-x-E(U)+b\hat i+c\hat j +d\hat k)\right\}. 
\end{equation}
We use the logistic kernel, $K(t)=1/(2+e^{-t}+e^t)$, for its entirety and ease in evaluation at tessarines. As shown in Appendix~\ref{app:logittess}, evaluating the logistic kernel at a tessarine $C=a+b\hat i+c\hat j+d\hat k$ gives  
\begin{equation}
    \label{eq:kerneleval}
    \begin{aligned}
    \Re\{K(C)\}  & 
    = \frac{2 \xi(a, b, c, d)}{\{\xi(a, b, c, d)\}^2+\{\psi(a, b, c, d)\}^2}, 
    \end{aligned}
\end{equation}
with $\xi(a, b, c, d)=2\{2+\cosh(a+c)\cos(b+d)+\cosh(a-c)\cos(b-d)\}$ and $\psi(a, b, c, d)=\sinh(a+c)\sin(b+d)+\sinh(a-c)\sin(b-d)$. Other kernel choices are possible provided that the kernel is defined over the entire tessarine ring. 

\cite{stefanski1990deconvolving} proposed to estimate $f(X)=K_h(X-x)$ via the deconvoluting kernel, $L_h(W-x)$, where $L_h(\cdot)=(1/h)L(\cdot/h)$, and 
\begin{equation}
    L(x) =\frac{1}{2\pi} \int e^{-itx}\frac{\phi_{K}(t)}{\phi_{U}(-t/h)}dt,
    \label{eq:kernelL}
\end{equation}
in which $\phi_{K}(\cdot)$ is the Fourier transform of  $K(\cdot)$, and $\phi_{U}(\cdot)$ is the characteristic function of  $U$. Under certain conditions to guarantee a well-defined integral in \eqref{eq:kernelL}, it has been shown that $E\{L_{h}(W-x)\mid X\}=K_{h}(X-x)$, and thus $L_{h}(W-x)$ is an unbiased estimator of $f(X)$. These conditions are imposed on $\phi_K(t)$ depending on the rate of convergence of $\phi_U(-t/h)$ towards zero as $h$ tends to zero. Using the deconvoluting kernel, one attains an estimator of $p_X(x)$, referred to as the \underline{d}econvoluting \underline{k}ernel density estimator, $\hat p_{\text{dk}}(x)=(1/n)\sum_{r=1}^n L_h(W_r-x)$. This estimator requires the knowledge of $\phi_U(\cdot)$, whereas the debiased estimator $\hat p_{\text{tc}}(x)$ in \eqref{eq:denest} only needs the first four moments of $U$. Moreover, computing $\hat p_{\text{dk}}(x)$ requires a great deal of caution when evaluating \eqref{eq:kernelL} with the untamed integrand. In comparison, using \eqref{eq:kerneleval}, computing $\hat p_{\text{tc}}(x)$ is a much simpler numerical task. As witnessed in our extensive simulation study presented in Section~\ref{sec:simu}, this simplicity dramatically reduces computation time at a small cost to accuracy. 

Besides the simulation study on  kernel density estimation discussed in this subsection, also included in the simulation experiments are the M-estimation in Poisson regression and logistic regression, which are analytically less transparent when it comes to comparing between different estimators than the comparisons in linear regression and polynomial regression.   

\section{Simulation Study}
\label{sec:simu}
\subsection{Experiments in Poisson regression}\label{sec:poissim}
Using simulated data from a Poisson regression model with error-in-covariate, we compare three estimators for regression coefficients $\Theta=(\beta_0, \beta_1)^\top$ using the maximum likelihood estimator based on error-free data as a reference estimator, denoted by $\hat \Theta$. The three estimators include the naive estimator based on error-contaminated data, denoted by $\hat \Theta_{\text{naive}}$, the corrected score estimator based on the score $\Psi_{\text{cs}}(Y, W, \Theta)$ in \eqref{eq:poiscs}, denoted by $\hat \Theta_{\text{cs}}$, and the tessarine-based estimator corresponding to the debiased score $\tilde \Psi(Y, W, \Theta)$ in \eqref{eq:poisdebias}, denoted by $\hat \Theta_{\text{tc}}$. 

Define $\lambda=\text{var}(X)/\{\text{var}(X)+\sigma_u^2\}$ as the reliability ratio \citep[][Section 3.2.1]{carroll2006measurement}. 
After generating a random sample of size $n \in \{250, 500\}$ from $N(0, 1)$ as the error-free covariate data $\{X_r\}_{r=1}^n$, we generate $\{(Y_r, W_r)\}_{r=1}^n$ according to \eqref{eq:poisreg}, with 
$\beta_{0}=1$ and $\beta_{1}=-1$, where $\{U_r\}_{r=1}^n$ is a random sample from a gamma distribution with the shape parameter equal to $4(1-\lambda)/\lambda$ and a scale parameter of 0.5, in which $\lambda \in \{0.8, 0.9, 1\}$. This design of measurement error distribution gives $\mu_3=\sigma_u^2=(1-\lambda)/\lambda$ and a kurtosis of $3[1+\{\lambda/(1-\lambda)\}^{1/2}]$. Moreover, under this design, real-valued solutions to \eqref{eq:findbcd} exist. At each simulation setting specified by the levels of $n$ and $\lambda$, we use 1000 Monte Carlo replicates to compare $\hat \Theta_{\text{naive}}$, $\hat \Theta_{\text{cs}}$, and $\hat \Theta_{\text{tc}}$. We acknowledge that, among the three estimators, only $\hat \Theta_{\text{tc}}$ is developed without assuming mean-zero measurement error. To make $\hat \Theta_{\text{naive}}$ and $\hat \Theta_{\text{cs}}$ relatively comparable with $\hat \Theta_{\text{tc}}$, we compute $\hat \Theta_{\text{naive}}$ and $\hat \Theta_{\text{cs}}$ based on covariate data contaminated by the centered measurement error $\{U_r-2(1-\lambda)/\lambda\}_{r=1}^n$. Similar treatments on data used by  competing methods are applied in all our empirical study. 

\begin{figure}[!h]
\centering
    \includegraphics[scale=.48]{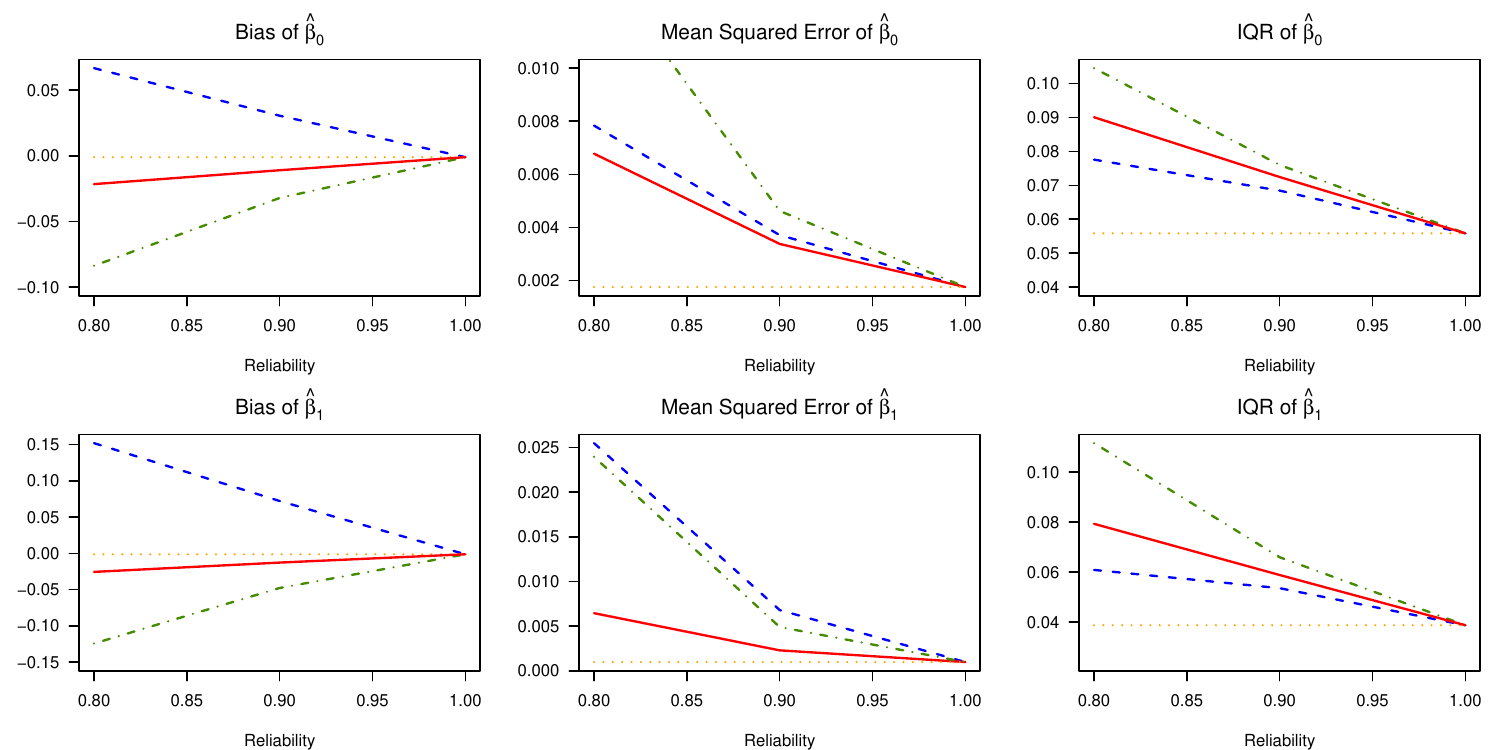}
  \caption{\label{fig:pois1}Empirical bias, mean squared errors, and interquartile ranges of four estimators for $\beta_0$ (top panels) and for $\beta_1$ (bottom panels) as the regression coefficients $\Theta=(\beta_0, \beta_1)^\top$ in the Poisson regression model when $n=250$: $\hat \Theta$ based on error-free data (dotted lines), $\hat \Theta_{\text{naive}}$ (dashed lines), $\hat \Theta_{\text{cs}}$ (dot-dashed lines), and $\hat \Theta_{\text{tc}}$ (solid lines). }
\end{figure}

Three metrics are used to assess the performance of an estimator. Take the slope parameter $\beta_1$ as an example and denote by $\hat \beta_{1,q}$ an estimate of it resulting from a considered estimation method applied to the $q$th Monte Carlo replicate dataset. Two of the metrics are the empirical bias defined as $(1000)^{-1}\sum_{q=1}^{1000}(\hat{\beta}_{1,q}-\beta_1)$, and 
the empirical mean squared error defined as $(1000)^{-1}\sum_{q=1}^{1000}(\hat{\beta}_{1,q}-\beta_1)^{2}$. A third metric is the empirical interquartile range of the estimates $\{\hat \beta_{1,q}\}_{q=1}^{1000}. $ Figure \ref{fig:pois1} depicts these metrics for $\hat \Theta$, $\hat \Theta_{\text{naive}}$, $\hat \Theta_{\text{cs}}$, and $\hat \Theta_{\text{tc}}$ as $\lambda$ varies when $n=250$. The counterpart results when $n=500$ are provided in the Supplementary Material. These graphical summaries show that, when bias and mean squared error are concerned, $\hat \Theta_{\text{tc}}$ is the closest to the reference estimator $\hat \Theta$ that is unattainable in the presence of measurement error. Although $\hat \Theta_{\text{cs}}$ slightly improves upon $\hat \Theta_{\text{naive}}$ in terms of bias, it is much less satisfactory than $\hat \Theta_{\text{tc}}$ and is subject to high variability. This illustrates the adverse effects of the departure from normality on the existing corrected score estimator. Admittedly, $\hat \Theta_{\text{tc}}$ tends to be more variable than $\hat \Theta_{\text{naive}}$, but the bias reduction achieved by $\hat \Theta_{\text{tc}}$  outweighs the inflation in variability, resulting in a much lower mean squared error than that of $\hat \Theta_{\text{naive}}$. 

\subsection{Experiments in logistic regression}\label{logsim}
In the context of logistic regression, we compare four estimators for the regression coefficients $\Theta=(\beta_0, \beta_1)^\top$, using the maximum likelihood estimator based on error-free data, $\hat \Theta$, as a reference estimator. The four considered estimators are the naive estimator $\hat \Theta_{\text{naive}}$, the Monte Carlo corrected score estimator $\hat \Theta_{\text{mc}}$ based on the limiting version of $\Psi_{\text{mc}}(Y, W, \Theta)$ in \eqref{eq:logitmccs} as $M\to \infty$ \citep[see equations (16),][]{StefNovick2002},  the conditional score estimator $\hat \Theta_{\text{cd}}$ based on the score $\Psi_{\text{cd}}(Y, W, \Theta)$ in \eqref{eq:logistcds}, and the estimator $\hat \Theta_{\text{tc}}$ based on the debiased score $\tilde \Psi_{\text{tc}}(Y, W, \Theta)$ in \eqref{eq:logisttc}. 

\begin{figure}[!h]
\centering
    \includegraphics[scale=.48]{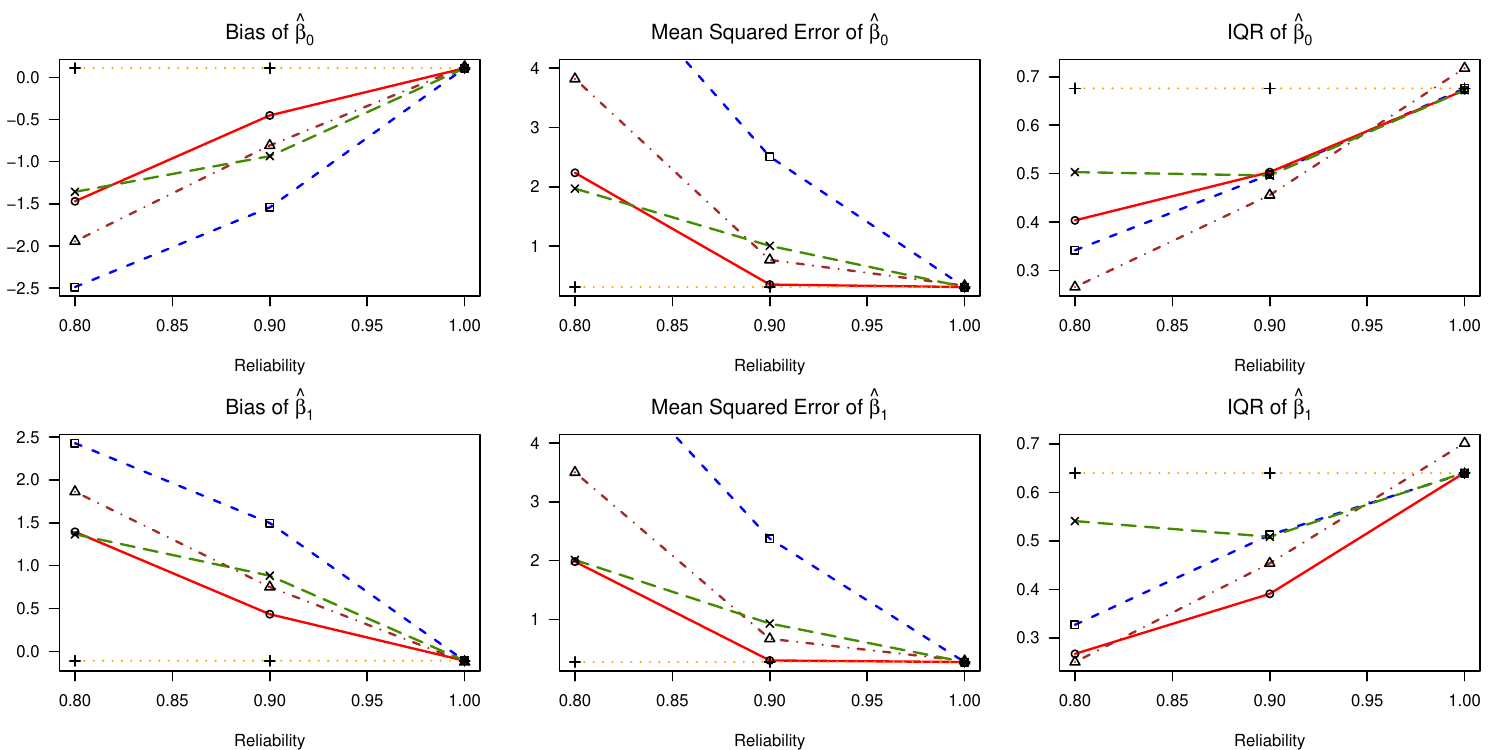}
    \caption{\label{fig:log1}Empirical bias, mean squared errors, and interquartile ranges of five estimators for $\beta_0$ (top panels) and for $\beta_1$ (bottom panels) as the regression coefficients $\Theta=(\beta_0, \beta_1)^\top$ in the logistic regression when $n=500$: $\hat \Theta$ (dotted lines, cross points), $\hat \Theta_{\text{naive}}$ (short dashed lines, square points),  $\hat \Theta_{\text{mc}}$ (dot-dashed lines, triangle points), $\hat \Theta_{\text{cd}}$ (long dashed lines, x points), and $\hat \Theta_{\text{tc}}$ (solid lines, circle points).}
\end{figure}

After generating error-free covariate data $\{X_r\}_{r=1}^n$ from $N(1, 1)$, we generate the binary responses and error-contaminated covariate data $\{(Y_r, W_r)\}_{r=1}^n$ according to \eqref{eq:logitreg}, with $\beta_{0}=5$, $\beta_{1}=-5$, and $\{U_r\}_{r=1}^n$ as described in Section \ref{sec:poissim}. At each simulation setting specified by $n\in \{500, 1000\}$ and $\lambda \in \{0.8, 0.9, 1\}$, we estimate $\Theta$ using the four estimators. Figure~\ref{fig:log1} presents the empirical bias, mean squared error, and interquartile range of each estimator across 1000 Monte Carlo replicates when $n=500$. The counterpart results when $n=1000$ are provided in the Supplementary Material. Even though the debiasing effect of the proposed estimator $\hat \Theta_{\text{tc}}$ cannot be justified via an infinite Taylor series due to the nonentirety of the expit function, such effect is evident in the actual estimate when compared with the naive estimate. The proposed estimator also substantially outperforms the other two non-naive estimators developed under the Gaussian measurement error assumption.   


\subsection{Experiments in density estimation}\label{kernelsim}
With the target of inferences being the probability density function of $X$, $p_X(x)$, we use the kernel density estimator based on error-free data $\{X_r\}_{r=1}^n$, $\hat  p_X(x)$, as a reference estimator to compare three density estimators based on error-contaminated data $\{W_r\}_{r=1}^n$. These include the naive estimator $\hat p_{\text{naive}}(x)$, the deconvoluting kernel density estimator $\hat p_{\text{dk}}(x)$, and the tessarine-based debiased estimator $\hat p_{\text{tc}}(x)$. In all considered estimators, the logistic kernel is used as $K(\cdot)$; and, for  $\hat p_{\text{dk}}(x)$, $\phi_U(\cdot)$ in \eqref{eq:kernelL} is taken as the Laplace characteristic function. 

At each of 500 Monte Carlo replicates at a simulation setting, the error-free data $\{X_r\}_{r=1}^n$ are generated from a two-component mixture distribution, $0.5 N(0, 1)+0.5N(5, 1)$, for $n\in \{50, 75\}$, following which we generate $\{W_r=X_r+U_r\}_{r=1}^n$, where $\{U_r\}_{r=1}^n$ is a random sample from a lognormal distribution with a logarithm of scale parameter $\sigma=\log[\{T(T-1)+1\}/2 T]$, with $T=2^{1/3}\sigma^{2}_u/[(\sigma^{2}_u[1+2\sigma^{2}_u]+\sigma_u^{2}(1+4\sigma^{2}_u)^{1/2}]^{1/3},$ and a logarithm of location $\mu=1/2(\log[\sigma^{2}/(\exp[\sigma^{2}-1])]),$ so that it yields $\mu_3=\sigma_u^2$ as $\lambda$ varies from 0.75 to 1 at increments of 0.025. Under this design, real-valued solutions to \eqref{eq:findbcd} do not exist.  
The metric used to assess the efficacy of a density estimator is the empirical mean integrated square error based on grid points $\{x_g\}_{g=1}^{100}$ equally spaced in $[-2, 8]$. For example,  let $\hat p_{\text{naive},q}(x)$ be the naive estimate from the $q$th Monte Carlo replicate, the empirical mean integrated square error of $\hat p_{\text{naive}}(x)$ is  $(500\times 100) ^{-1}\sum_{q=1}^{500} \sum_{g=1}^{100} \{ \hat p_{\text{naive},q}(x_g)-p_X(x_g)\}^2$. 

Bandwidth selection is an integral part of kernel density estimation. This problem in the error-free case has been well-studied \citep{hastie2009kernel}. In cases with error-contaminated data, \cite{delaiglebandwidth} provided several strategies for choosing the bandwidth in $\hat p_{\text{dk}}(x)$. These methods are tailored for estimation based on the deconvoluting kernel that involves a Fourier transform, and thus cannot be easily revised to select a bandwidth for $\hat p_{\text{tc}}(x)$. Deriving an objective function, such as the mean integrated squared error, that is practically useful for selecting a bandwidth in $\hat p_{\text{tc}}(x)$  is much less straightforward. Further, such an objective function depends on higher-order derivatives of $p_X(x)$ that need to be estimated, preferably using a similar strategy that leads to $\hat p_{\text{tc}}(x)$, which are topics beyond the scope of the current study. 
To empirically compare different density estimators here, we choose the bandwidth for each method that minimizes the corresponding empirical mean integrated squared error. To choose a bandwidth in the naive estimator $\hat p_{\text{naive}}(x)$, the true density of $W$ is used in place of $p_X(\cdot)$ in the empirical mean integrated squared error so that the method is still considered naive.

\begin{figure}[!h]
\centering
    \includegraphics[scale=.48]{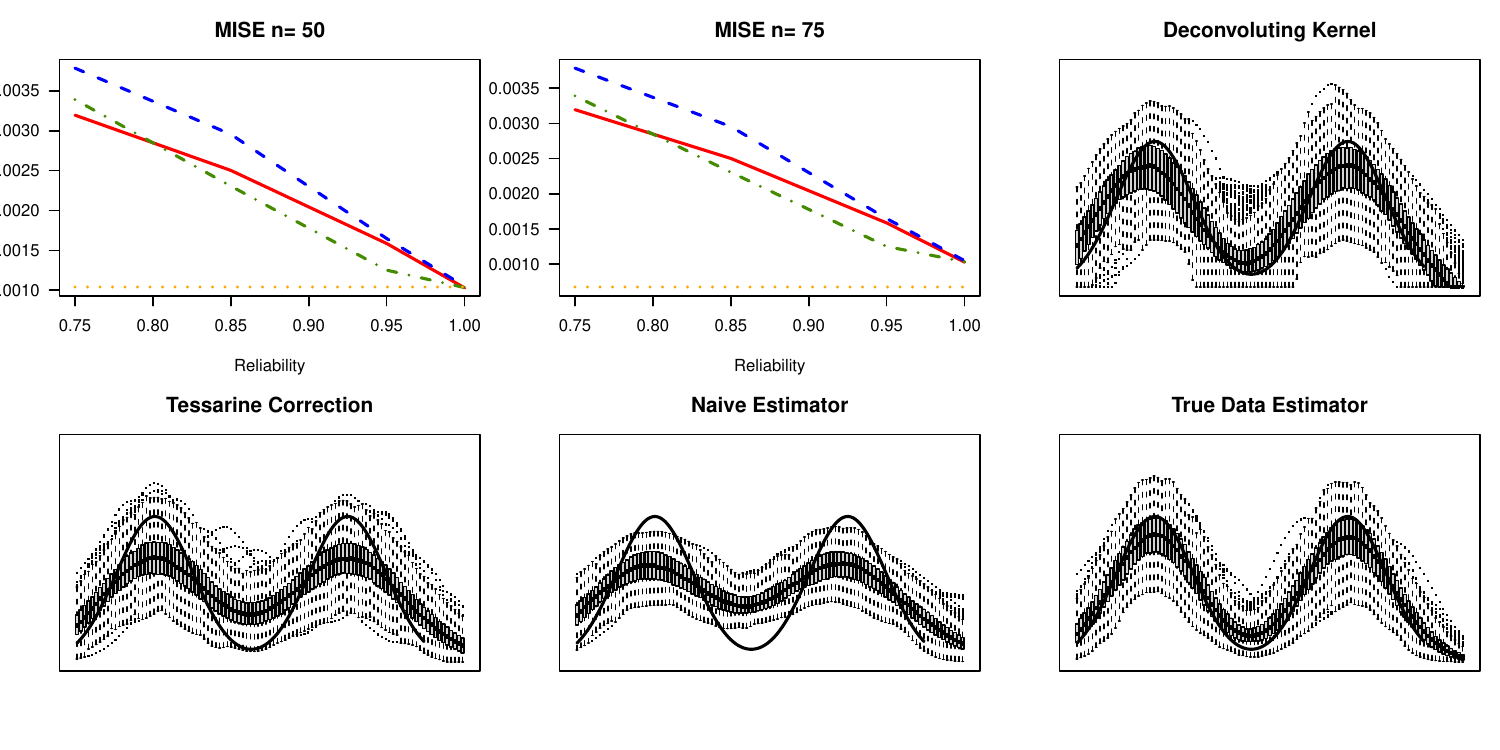}
    \caption{\label{fig:kde} Empirical mean integrated square errors of four density estimators when $n=50$ (top left panel) and $n=75$ (top middle panel): $\hat p_X(x)$ (dotted lines), $\hat p_{\text{naive}}(x)$ (dashed lines), $\hat p_{\text{dk}}(x)$ (dot-dashed lines), and $\hat p_{\text{tc}}(x)$ (solid lines), along with boxplots of estimates from each method when $n=50$ and $\lambda=0.9$ superimposed on the true density (solid lines).}
\end{figure}

Figure \ref{fig:kde} presents the empirical mean integrated error for the three considered estimators compared with the reference estimator $\hat p_X(x)$.  Even though the deconvoluting kernel density estimator $\hat p_{\text{dk}}(x)$ outperforms the debiased estimator $\hat p_{\text{tc}}(x)$ in terms of the mean integrated squared error, $\hat p_{\text{tc}}(x)$ significantly improves upon the naive estimator $\hat p_{\text{naive}}(x)$ and is less variable than $\hat p_{\text{dk}}(x)$ despite the lack of real-valued solutions to  \eqref{eq:findbcd}. Moreover, the computation time of  $\hat p_{\text{tc}}(x)$ is dramatically shorter than that of  $\hat p_{\text{dk}}(x)$. At a sample size of $n=50$,  it takes around 9 seconds to compute $\hat p_{\text{dk}}(x)$, whereas $\hat p_{\text{tc}}(x)$ takes 0.11 seconds to compute on a computer with an Intel Core i5 12600KF running at 4500 Mhz. 

 \subsection{Sensitivity analysis}
 Because the proposed tessarine-based debiased estimator depends on the first four moments of $U$, a legitimate question is the impact of moment misspecification on the proposed estimator. To empirically inspect the influence of moment misspecification on different estimators, we conduct a sensitivity analysis on the estimation of regression coefficients in a Poisson regression model in a simulation setting similar to that in Section~\ref{sec:poissim}.   

In particular, we generate a dataset of size $n=250$, $\{(X_r, Y_r, W_r)\}_{r=1}^n$, following the data generating mechanism described for $(X, Y, W)$ in Section~\ref{sec:poissim}, where $U$ follows a gamma distribution with the shape parameter equal to 4/9 and the scale parameter equal to 0.5, gamma(4/9, 0.5), resulting in $\lambda=0.9$. Then, instead of using the first four moments of gamma(4/9, 0.5) to find $(b, c, d)$ in the tessarine-based estimator $\hat \Theta_{\text{tc}}$, we assume the second, third, and fourth central moments of $U$ equal to $\mathcal{E}_1\sigma^2_u$, $\mathcal{E}_2\mu_3$, and $\mathcal{E}_3\mu_4$, respectively, 
where $\sigma_u^2$, $\mu_3$, and $\mu_4$ are the true moments of gamma(4/9, 0.5), $\mathcal{E}_1$, $\mathcal{E}_2$, and $\mathcal{E}_3$ are random numbers generated independently from a uniform distribution supported on $[1-\delta, \, 1+\delta]$, with $\delta$ varying from 0 to 0.6 at increments of 0.02. At each level of $\delta$, we repeatedly compute $\hat \Theta_{\text{tc}}$ under the wrong (when $\delta>0)$ assumption of the measurement error moments using 5,000 Monte Carlo replicates. For comparison, we also compute the naive estimate $\hat \Theta_{\text{naive}}$, which is not affected by $\delta$, and the corrected score estimate $\hat \Theta_{\text{cs}}$, which depends on the variance of $U$, according to \eqref{eq:poiscs}, and thus is affected by $\delta$. 

\begin{figure}[h]
        \centering
        \includegraphics[scale=.48]{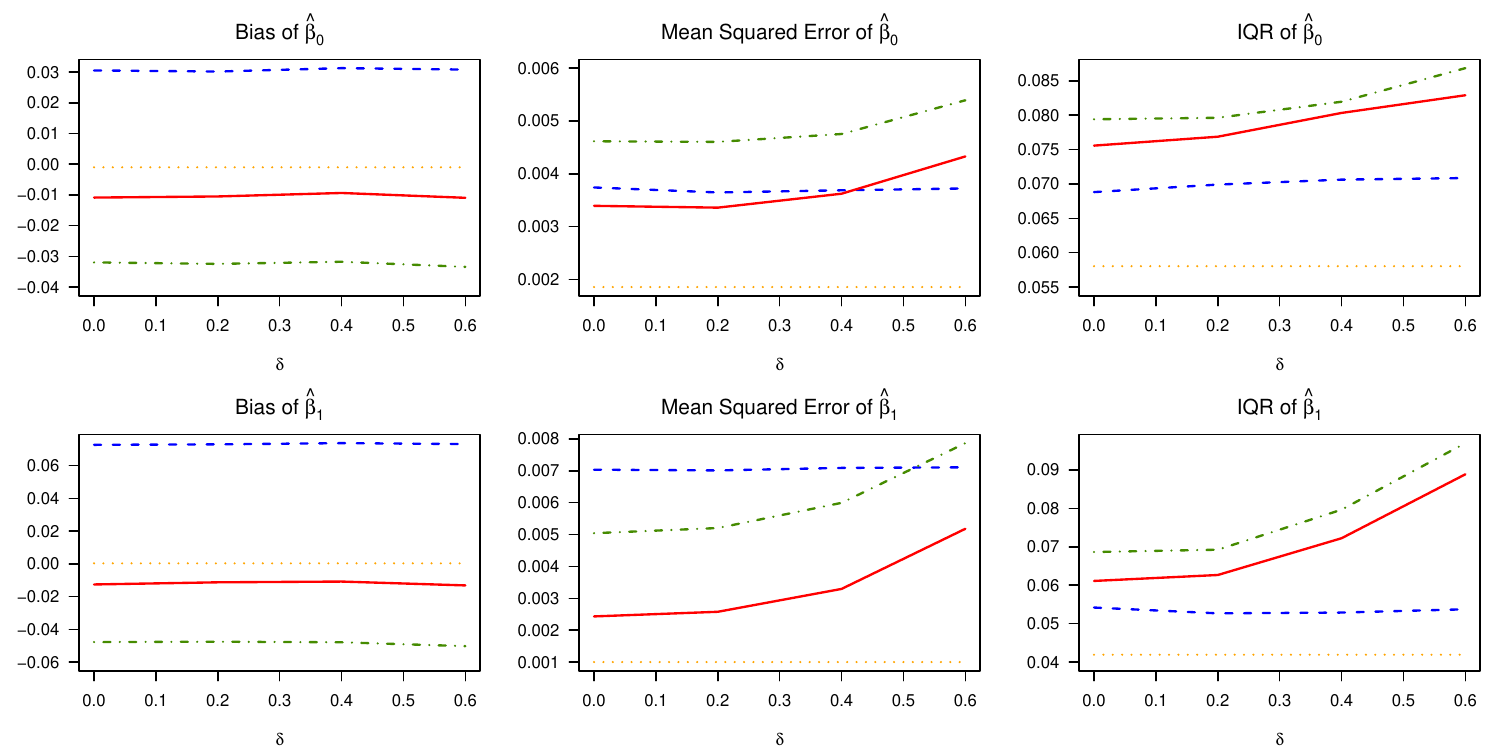}
        \caption{\label{fig:misspec}
        Empirical bias, mean squared errors, and interquartile ranges for $\beta_0$ (top panels) and for $\beta_1$ (bottom panels) as the regression coefficients $\Theta=(\beta_0, \beta_1)^\top$ in a Poisson regression model across 5,000 Monte Carlo replicates in the sensitivity analysis: $\hat \Theta$ (dotted lines), $\hat \Theta_{\text{naive}}$ (dashed lines), $\hat \Theta_{\text{cs}}$ (dot-dashed lines), and $\hat \Theta_{\text{tc}}$ (solid lines).} 
    \end{figure}

Figure~\ref{fig:misspec} provides the empirical bias, mean squared errors, and interquartile ranges of the four considered sets of estimates across 5,000 Monte Carlo replicates. Interestingly $\hat{\Theta}_{\text{tc}}$ and $\hat{\Theta}_{\text{cs}}$ are similar in the rate of deterioration as moments misspecification becomes more severe, despite the fact that $\hat{\Theta}_{\text{tc}}$ involves more misspecified moments that   $\hat{\Theta}_{\text{cs}}$ does. Even though $\hat \Theta_{\text{tc}}$ is compromised by the misspecification, its performance remains competitive over a wide range of $\delta$, that is, even when the misspecification is moderate. We thus conclude that, in the presence of non-Gaussian measurement error, even when the first four moments of $U$ are estimated and thus subject to misspecification, the tessarine-based debiased estimator can still outperform the naive estimator, and is more preferable than the corrected score estimator developed under the assumption of Gaussian measurement error.     

\section{Application in  hockey data}
\label{sec:real}
An important metric used to assess the effectiveness of hockey players in the National Hockey League is the expected goals. Sports analysts use factors such as shot quality, shot location, and speed to calculate the number of expected goals a player would have scored on average during a game. This metric has a natural equivalent for the goaltender (as the player attempting to stop shots), which is the expected goals allowed. In this study, we consider a Poisson model in \eqref{eq:poisreg} for the number of wins for each goaltender ($Y$), regressing on the number of goals allowed ($X$), based on a dataset consisting of $n=88$ records of win data from Hockey Reference (\url{https://www.hockey-reference.com/leagues/NHL_2022_goalies.html}) and expected goals data from Moneypuck.com (\url{https://moneypuck.com/data.htm}). 

\begin{figure}[h]\centering
    \includegraphics[scale=.48]{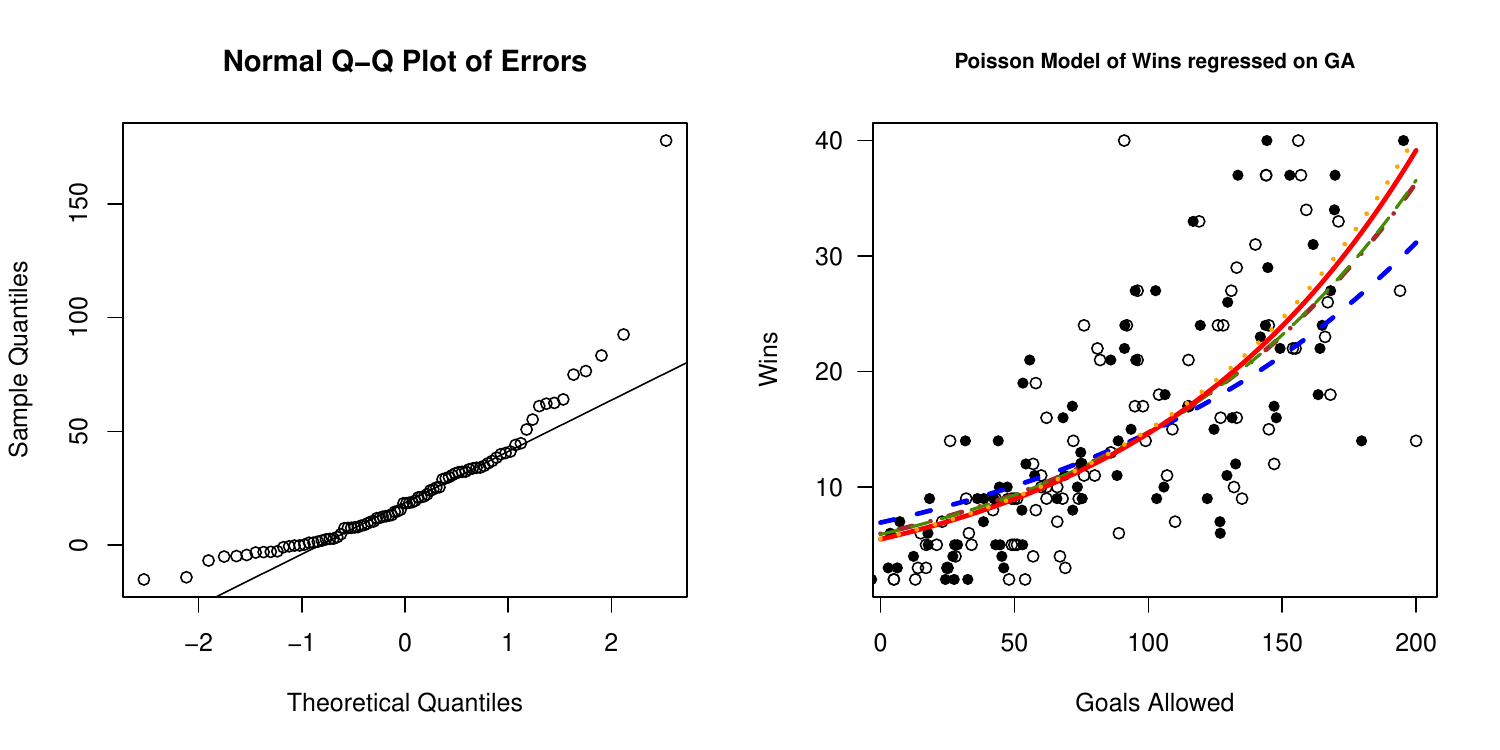}
\caption{\label{fig:goalies}The normal quantile-quantile plot (left panel) for expected goals$-$actual goals, and five estimates for the regression function (right panel): $\hat m(x)$ (dotted line), $\hat m_{\text{naive}}(x)$ (dashed line), $\hat m_{\text{cs}}(x)$ (long dashed line), $\hat m_{\text{cc}}(x)$ (dot-dashed line), and $\hat m_{\text{tc}}(x)$ (solid line), superimposed on error-free data (empty circles) and error-contaminated data (solid circles). }
\end{figure}

To demonstrate inferential methods for error-in-variable problems, we create a surrogate of $X$ by adding random noise from $\text{gamma}(1, 20)$ to the expected goals recorded in this dataset. The so-obtained surrogate covariate $W$ corresponds to the measurement error $U=W-X$, with an estimated skewness of 2.3 and an estimated kurtosis of 12.06. Figure~\ref{fig:goalies} presents in the left panel the normal quantile-quantile plot of the measurement errors. This plot, along with the estimated skewness and kurtosis, clearly suggests a deviation from normality, especially in the tails that appear to be heavier than that for Gaussian errors. 
These observations warrant regression analysis which accounts for non-Gaussian measurement errors that may not have mean zero. 

Figure~\ref{fig:goalies} shows in the right panel five estimates for the regression function $m(x)=E(Y|X=x)=\exp(\beta_0+\beta_1 x)$, including the estimate based on the maximum likelihood estimate of $\Theta=(\beta_0, \beta_1)^\top$ using error-free data $\{(Y_r, X_r)\}_{r=1}^{88}$, denoted by $\hat m(x)$, its naive counterpart based on the error-contaminated data $\{(Y_r, W_r)\}_{r=1}^{88}$, denoted by $\hat m_{\text{naive}}(x)$, the corrected score estimate $\hat m_{\text{cs}}(x)$, the estimate involving complex constants in the spirit of $\hat f_{\text{cc}}(x)$, denoted by $\hat m_{\text{cc}}(x)$, and the tessarine-based debiased estimate $\hat m_{\text{tc}}(x)$. Evidently, $\hat m_{\text{tc}}(x)$ resembles $\hat m(x)$ most closely among the four estimated regression functions based on error-contaminated data. Both $\hat m_{\text{naive}}(x)$ and $\hat m_{\text{cs}}(x)$ seem to suggest a much weaker association between the number of wins and the number of goals allowed than what $\hat m(x)$ indicates, and $\hat m_{\text{cc}}(x)$ is similar to $\hat m_{\text{cs}}(x)$. Table~\ref{t:hockey} provides the estimates for $\Theta$ corresponding to the five estimated regression functions, along with the estimated standard errors based on the sandwich variance estimator \citep[][Section 7.2]{boos2013essential}. All estimated covariate effects based on error-contaminated data are attenuated when compared to the estimate based on error-free data. But the debiased estimate $\hat \beta_{1,\text{tc}}(x)$ is least attenuated, and less variable than the other two non-naive estimations using the same data.

\begin{table}[h]
\centering
\caption{\label{t:hockey}Regression coefficients estimates in the estimated regression functions based on the hockey data, with the corresponding estimated standard errors in parentheses.}
\begin{tabular}{ccc}
\hline
Regression function & $\hat{\beta}_{0}$ & $\hat{\beta}_{1}$ \\ \hline 
$\hat m(x)$ & 1.7089  (0.0687)  & 0.0100  (0.0006)   \\ 
$\hat m_{\text{naive}}(x)$  & 1.9350  (0.0659)  & 0.0075 (0.0005) \\ 
$\hat m_{\text{cs}}(x)$  & 1.7747 (0.2092)  & 0.0091 (0.0022) \\ 
$\hat m_{\text{cc}}(x)$  & 1.7842 (0.2052)  & 0.0090 (0.0021) \\
$\hat m_{\text{tc}}(x)$  & 1.7008 (0.0921)  & 0.0098 (0.0010) \\
\hline
\end{tabular}
\end{table}

\section{Discussion}
\label{sec:disc}
A tessarine is a four-dimensional hypercomplex number. To continue on the path of bias reduction achieved by the proposed tessarine-based estimator $\hat f_{\text{tc}}(x)$, one can extend the debiased estimator construction based on a higher dimensional hypercomplex number, such as octonions that are eight-dimensional \citep{cayley1845xxviii}. However,  algebraic manipulations quickly become more complicated when using higher dimensional hypercomplex numbers in constructing an estimator, and the bias reduction beyond what is accomplished by $\hat f_{\text{tc}}(x)$ is expected to be lessened. This is because,  under certain assumptions, including the entirety of $f(x)$ and moment conditions on $U$, higher-order bias in \eqref{eq:properCbias}--\eqref{eq:naivebias} is dominated by terms of lower orders. 

Except for the example of linear regression in Section~\ref{sec:linearreg}, we have focused on error-in-variable problems with a scalar error-prone variable, allowing us to focus on methodology development mostly involving univariate hypercomplex numbers. Generalization of this line of development adapted to more general error-in-variables problems with multiple variables subject to correlated non-Gaussian measurement error contamination is an interesting topic for future research.


\renewcommand\thesection{Appendix \arabic{section}}
\begin{appendix}
\renewcommand*{\theequation}{A.\arabic{equation}}
\setcounter{equation}{0}
\section{Proof of Lemma~\ref{lem:zeromom}}
\label{app:provelem1}
\begin{proof}
By the definition of circular symmetry, $e^{it}C$ and the circularly symmetric $C$ follow the same distribution for an arbitrary constant $t$ on the real line $\mathbb{R}$. Hence, assuming moments of $C$ of all orders exist, 
$$E(e^{i \ell t}C^\ell)=E(C^\ell), \text{ for all $t \in \mathbb{R}$ and $\ell=1, 2, \ldots$},$$
which is equivalent to  
$$E(C^\ell)(1-e^{i\ell t})=0 , \text{ for all $t \in \mathbb{R}$ and $\ell=1, 2, \ldots$}.$$
By the arbitrariness of $t$, the above identity holds if and only if $E(C^\ell)=0$, for $\ell=1, 2, \ldots$. This completes the proof. 
\end{proof}

\section{Proof of Theorem~\ref{thm:proper}}
\label{app:provethm2.2}
\begin{proof}
    With $C=U-E(U)+i\{V-E(V)\}$, $E(C)= E\{U-E(U)\}+i E\{V-E(V)\}=0$. Hence, the  pseudovariance is $\text{pvar}(C)=E(C^2)$. Because 
    $$C^2=\{U-E(U)\}^2-\{V-E(V)\}^2+2i \{U-E(U)\}\{V-E(V)\},$$ we have 
$\text{pvar}(C) = \text{var}(U)-\text{var}(V)+2i E\{U-E(U)\}E\{V-E(V)\} = 0$, since $U$ and $V$ are independent and identically distributed, thus have the same variance. Because $C$ has a vanishing pseudovariance, by definition, $C$ is proper. This completes the proof. 
\end{proof}

\section{Proof of Lemma~\ref{lem:power}}
\label{app:provelem2}
\begin{proof}
Diagonalizing the matrix-version $C$ yields
$$C = \begin{bmatrix} 
-\displaystyle{\frac{1}{\sqrt{2}}} && \displaystyle{\frac{1}{\sqrt{2}}}\\ 
\displaystyle{\frac{1}{\sqrt{2}}} && \displaystyle{\frac{1}{\sqrt{2}}}
\end{bmatrix} 
\begin{bmatrix} 
a-c+i(b-d) & 0 \\ 
0 & a+c+i(b+d) 
\end{bmatrix}
\begin{bmatrix} 
-\displaystyle{\frac{1}{\sqrt{2}}}&& \displaystyle{\frac{1}{\sqrt{2}}}\\
\displaystyle{\frac{1}{\sqrt{2}}} && \displaystyle{\frac{1}{\sqrt{2}}}
\end{bmatrix}.$$
Because $(-1/\sqrt{2}, 1/\sqrt{2})^\top$ and $(1/\sqrt{2}, 1/\sqrt{2})^\top$ are orthonormal, we have
\begin{align*}
        &\ C^{\ell} \\
       = &\
        \begin{bmatrix} 
        -\displaystyle{\frac{1}{\sqrt{2}}} && \displaystyle{\frac{1}{\sqrt{2}}}\\ 
        \displaystyle{\frac{1}{\sqrt{2}}} && \displaystyle{\frac{1}{\sqrt{2}}}
        \end{bmatrix} 
        \begin{bmatrix} 
        \{a-c+i(b-d)\}^\ell & 0 \\
        0 & \{a+c+i(b+d)\}^\ell 
        \end{bmatrix}
        \begin{bmatrix} 
        -\displaystyle{\frac{1}{\sqrt{2}}} && \displaystyle{\frac{1}{\sqrt{2}}}\\ 
        \displaystyle{\frac{1}{\sqrt{2}}} && \displaystyle{\frac{1}{\sqrt{2}}}
        \end{bmatrix}\\
     =   &\ \frac{1}{2}\left(\left[\{a-c+i(b-d)\}^\ell+\{a+c+i(b+d)\}^\ell\right]\bI_2\right. \\
         &\ \left. + \left[\{a+c+i(b+d)\}^\ell-\{a-c+i(b-d)\}^\ell \right] \bJ_2\right).
\end{align*}
The diagonal entry of $C^\ell$ is thus, by Euler's formula, 
\begin{align*}
    &\ \frac{1}{2} \left(\left\{(a-c)^2+(b-d)^2\right\}^{\ell/2}\exp[i\ell \times \text{arg}\{a-c+i(b-d)\}]\right. \\
    &\ \left. +\left\{(a+c)^2+(b+d)^2\right\}^{\ell/2}\exp[i\ell\times \text{arg}
    \{a+c+i(b+d)\}]\right),
\end{align*}
where $\text{arg}(t)$ is the  argument of a complex number $t$. Extracting the real part of the above expression gives  
\begin{align*}
     \Re(C^\ell) =&\ \frac{1}{2}\left[\left\{(a-c)^{2}+(b-d)^{2}\right\}^{\ell/2}\cos(\ell\times \text{arg}\{a-c+i(b-d)\})\right.\\
     &\ \left.+\left\{(a+c)^{2}+(b+d)^{2}\right\}^{\ell/2}\cos(\ell \times\text{arg}\{a+c+i(b+d)\})\right].    \end{align*}
This completes the proof.
\end{proof}

\renewcommand*{\theequation}{D.\arabic{equation}}
\setcounter{equation}{0}
\section{Proof of Theorem~\ref{thm:tesexp}}
\label{app:provethm3.1}
\begin{proof}
   For a square real or complex matrix $C$, the matrix exponential is $\exp(C)=\sum_{\ell=0}^\infty  C^\ell/{\ell !}$ (\url{https://en.wikipedia.org/wiki/Matrix_exponential}).
   
   Expressing the tessarine $C=a+b \hat{i}+c\hat{j}+d\hat{k}$ as a $2 \times 2$ complex matrix and using Lemma~\ref{lem:power}, we have 
     \begin{align}\label{matexp}
         \exp(C) & =\frac{1}{2}
         \begin{bmatrix}
         \mathcal{A} && \mathcal{B} \\ 
         \mathcal{B} && \mathcal{A}
         \end{bmatrix}, 
     \end{align}
     where 
     \begin{align*}
         \mathcal{A} & = \sum_{\ell=0}^\infty 
         \frac{\{a-c+i(b-d)\}^\ell}{\ell!}+\sum_{\ell=0}^\infty \frac{\{a+c+i(b+d)\}^\ell}{\ell!} \\
         & = \exp\{a-c+i(b-d)\}+\exp\{a+c+i(b+d)\}   \\
         & =2 \exp(a+ib)\cosh(c+id),
         \end{align*}
and 
    \begin{align*}
         \mathcal{B} & = \sum_{\ell=0}^\infty 
         \frac{\{a+c+i(b+d)\}^\ell}{\ell!}-\sum_{\ell=0}^\infty \frac{\{a-c+i(b-d)\}^\ell}{\ell!} \\
         & = \exp\{a+c+i(b+d)\}-\exp\{a-c+i(b-d)\}   \\
         & =2 \exp(a+ib)\sinh(c+id).
         \end{align*}
Therefore, 
    \begin{align*}    
    \exp(C) & = \exp(a+bi)
    \begin{bmatrix} 
    \cosh(c+di) && \sinh(c+di) \\ \sinh(c+di) && \cosh(c+di)
    \end{bmatrix}\\
    & = \exp(a+bi)\{\cosh(c+di)\bI_2+\sinh(c+di)\bJ_2\}.
    \end{align*}
   Extracting the real part of the diagonal entry of $\exp(C)$ gives 
   \begin{align*}
       \Re\{\exp(C)\}& = \Re\{\exp(a+bi)\cosh(c+di)\} \\
       & = \exp(a)\{\cos(b)\cos(d)\cosh(c)-\sin(b)\sin(d)\sinh(c)\}.
   \end{align*}
This completes the proof.
\end{proof}

\section{Derivations of the real parts in the debiased score in logistic regression}
\label{app:logistic}

The debiased score $\tilde \Psi(Y, W, \Theta)$ involves $\Re\{\mathcal{G}(\beta_0+\beta_1\mathscr{T})\}$ and $\Re\{\mathcal{G}(\beta_0+\beta_1\mathscr{T})\mathscr{T}\}$, where $\mathcal{G}(t)=1/\{1+\exp(-t)\}$ and $\mathscr{T}=W-E(U)+b\hat i +c\hat j +d \hat k$. To derive these real parts, we first derive the matrix version of the expit function evaluated at a tessarine $t=R_t+I_t \hat i +J_t \hat j +K_t \hat k$, where $R_t, I_t, J_t, K_t \in \mathbb{R}$ are the real part and the imaginary parts of $t$.

Firstly, the matrix form of the tessarine basis element 1 is $\mathbbm{1}=\bI_2$, that is, the $2\times 2$ identity matrix. Secondly, by Theorem~\ref{thm:tesexp}, 
\begin{align*}
    \exp(-t) & = \exp(-R_t-I_t i) 
    \begin{bmatrix}
        \cosh(-J_t-K_t i) && \sinh(-J_t-K_t i) \\
      \sinh(-J_t-K_t i)  && \cosh(-J_t-K_t i) 
    \end{bmatrix}\\
    & = \exp(-R_t-I_t i) 
    \begin{bmatrix}
        \cosh(J_t+K_t i) && -\sinh(J_t+K_t i) \\
      -\sinh(J_t+K_t i)  && \cosh(J_t+K_t i) 
    \end{bmatrix}.
\end{align*}
Summing these two matrices gives 
\begin{align*}
    &\ D(t) = \mathbbm{1}+\exp(-t) \\
   = &\ 
    \begin{bmatrix}
        1+ \exp(-R_t-I_t i)\cosh(J_t+K_t i) && -\exp(-R_t-I_t i)\sinh(J_t+K_t i) \\
       -\exp(-R_t-I_t i)\sinh(J_t+K_t i) && 1+ \exp(-R_t-I_t i)\cosh(J_t+K_t i)
    \end{bmatrix}.
\end{align*}
Thirdly, expressing the reciprocal of a tessarine as the inverse of a complex matrix, we have $\mathcal{G}(t)=\{D(t)\}^{-1}$. Lastly, $\Re\{\mathcal{G}(t)\}$ is equal to the real part of the first diagonal entry of $\mathcal{G}(t)$. Given $D(t)$ above, we have the first diagonal entry of $\mathcal{G}(t)$ as follows,  
\begin{align*}
    &\ \mathcal{G}(t)[1, 1] =
    \frac{D(t)[2, 2]}{\det\{D(t)\}} \\
    = &\ \frac{1+\exp(-R_t-I_t i)\cosh(J_t+K_t i)}{\{1+\exp(-R_t-I_t i)\cosh(J_t+K_t i)\}^2-\{\exp(-R_t-I_t i)\sinh(J_t+K_t i)\}^2}.
\end{align*}
Extracting the real part of $\mathcal{G}(t)[1, 1]$ gives $\Re\{\mathcal{G}(t)\}$. Setting $t=\beta_0+\beta_1 \mathscr{T}$ gives what we need as $\Re\{\mathcal{G}(\beta_0+\beta_1\mathscr{T})\}$ in the debiased score. 

Consider another tessarine $s=R_s+I_s \hat i +J_s \hat j +K_s \hat k$. Then 
\begin{align*}
    \mathcal{G}(t)s & = \{D(t)\}^{-1} 
    \begin{bmatrix}
      R_s+I_s i && J_s+K_s i \\
      J_s+K_s i  && R_s+I_s i
    \end{bmatrix},
\end{align*}
and its first diagonal entry is

\noindent
\begin{align*}
   &\ \{\mathcal{G}(t)s\}[1,1]\\
   = &\ \frac{D(t)[2, 2] (R_s+I_si)-D(t)[1, 2](J_s+K_si)}{\det\{D(t)\}}\\
   = &\ \frac{\{1+\exp(-R_t-I_t i)\cosh(J_t+K_t i)\} (R_s+I_s  i)+\exp(-R_t-I_t i)\sinh(J_t+K_t i)(J_s+K_s i)}{\{1+\exp(-R_t-I_t i)\cosh(J_t+K_t i)\}^2-\{\exp(-R_t-I_t i)\sinh(J_t+K_t i)\}^2}.
\end{align*}
Extracting the real part of the above expression gives $\Re(\mathcal{G}(t)s\}[1,1])$, and setting $t=\beta_0+\beta_1 \mathscr{T}$ and $s=\mathscr{T}$ gives what we need as $\Re\{\mathcal{G}(\beta_0+\beta_1\mathscr{T})\mathscr{T}\}$ in the debiased score. 

\section{Evaluation of the logistic kernel at a tessarine}
\label{app:logittess}
For $C=a\mathbbm{1}+b \mathbb{I}+c\mathbb{J}+d\mathbb{K}$, by Theorem~\ref{thm:tesexp}, we have 
\begin{align*}
    &\ 2+ \exp(C)+\exp(-C) \\ 
    = &\ 2 \bI_2+\exp(a+bi)\{\cosh(c+di)\bI_2+\sinh(c+di)\bJ_2\}\\
    &\ +\exp(-a-bi)\{\cosh(c+di)\bI_2-\sinh(c+di)\bJ_2\} \\
   = &\ 2 \{1+\cosh(a+bi)\cosh(c+di)\}\bI_2+2 \sinh(a+bi)\sinh(c+di)\bJ_2.
\end{align*}
Expressing $K(C)=1/(2+e^C+e^{-C})$ as the inverse of the above $2\times 2$ complex matrix, and using the result on the inverse of the sum of two matrices in \cite{miller1981inverse}, we have 
\begin{align*}
    &\ K(C) \\
   = &\ \frac{1}{2\{1+\cosh(a+bi)\cosh(c+di)\}}\left\{\bI_2-\frac{\sinh(a+bi)\sinh(c+di)}{1+\cosh(a+bi)\cosh(c+di)}\bJ_2\right\}. 
\end{align*}
It follows that 
\begin{align*}
    \Re\{K(C)\} & = \Re\{K(C)[1,1]\} \\
    & = \Re\left[ \frac{1}{2\{1+\cosh(a+bi)\cosh(c+di)\}}\right]\\
    & = \frac{2 \xi(a, b, c, d)}{\xi(a, b, c, d)^2+\psi(a, b, c, d)^2}, 
\end{align*}
with $\xi(a, b, c, d)=2\{2+\cosh(a+c)\cos(b+d)+\cosh(a-c)\cos(b-d)\}$ and $\psi(a, b, c, d)=\sinh(a+c)\sin(b+d)+\sinh(a-c)\sin(b-d)$.

\renewcommand*{\theequation}{G.\arabic{equation}}
\setcounter{equation}{0}
\renewcommand*{\thesubsection}{G.\arabic{subsection}}
\renewcommand*{\thefigure}{G.\arabic{figure}}
\setcounter{figure}{0}
 \section{Additional simulation results}
 \label{app:moresim}
\begin{figure}[!h]
\centering
    \includegraphics[scale=.48]{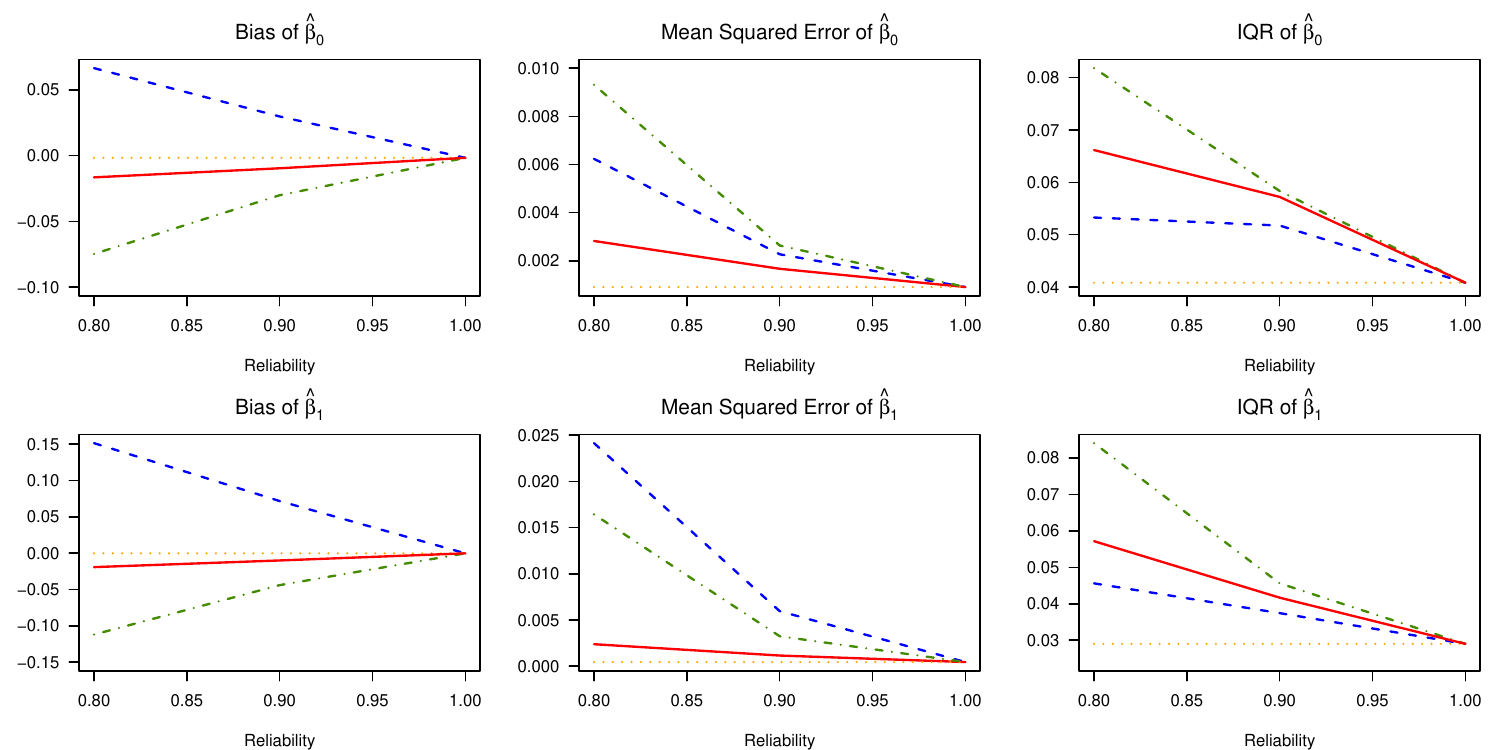}    \caption{\label{fig:pois2}Empirical bias, mean squared errors, and interquartile ranges of four estimators for $\beta_0$ (top panels) and for $\beta_1$ (bottom panels) as the regression coefficients $\Theta=(\beta_0, \beta_1)^\top$ in the Poisson regression model when $n=500$: the maximum likelihood estimator $\hat \Theta$ based on error-free data (dotted lines), the naive estimator $\hat \Theta_{\text{naive}}$ (dashed lines), the corrected score estimator $\hat \Theta_{\text{cs}}$ (dashed lines), and the tessarine-based estimator $\hat \Theta_{\text{tc}}$ (solid lines).}
\end{figure}

\begin{figure}[!h]
\centering
    \includegraphics[scale=.48]{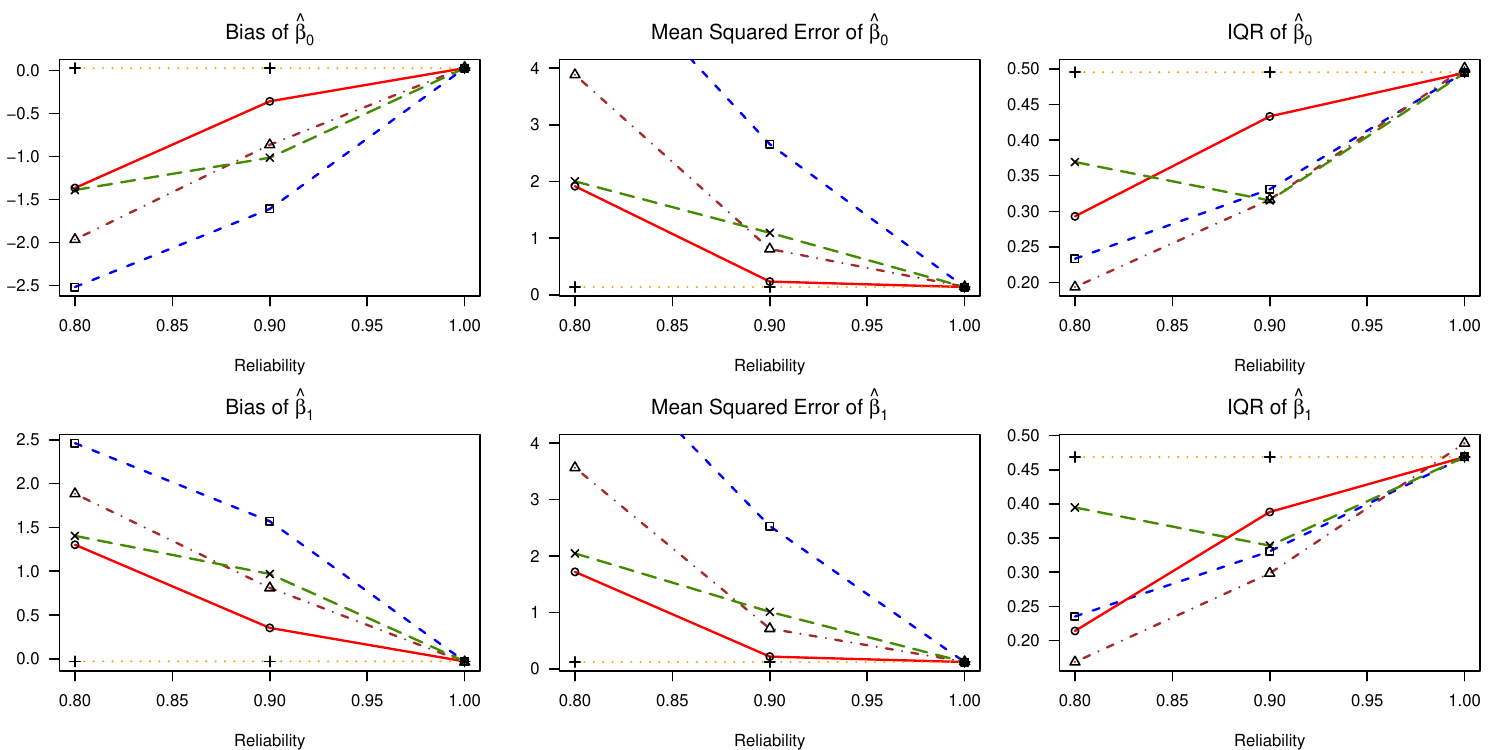}
    \caption{\label{fig:log2}Empirical bias, mean squared errors, and interquartile ranges of five estimators for $\beta_0$ (top panels) and for $\beta_1$ (bottom panels) as the regression coefficients $\Theta=(\beta_0, \beta_1)^\top$ in the logistic regression when $n=1000$: the maximum likelihood estimator $\hat \Theta$ based on error-free data (dotted lines, cross points), the naive estimator $\hat \Theta_{\text{naive}}$ (short dashed lines, square points), the Monte Carlo corrected score estimator $\hat \Theta_{\text{mc}}$ (dot-dashed lines, triangle points), the conditional score estimator $\hat \Theta_{\text{cd}}$ (long dashed lines, x points), and the tessarine-based estimator $\hat \Theta_{\text{tc}}$ (solid lines, circle points).}
\end{figure}

Figure~\ref{fig:pois2} is the counterpart of Figure~\ref{fig:pois1} with a larger sample size at $n=500$ in the context of Poisson regression. Figure~\ref{fig:log2} is parallel to Figure~\ref{fig:log1} with a large sample size of $n=1000$ in the context of logistic regression. 

\end{appendix}



\bibliographystyle{apalike}
\bibliography{refs.bib}       

\end{document}